\pdfoutput=1

\documentclass[aip,jcp,reprint,groupedaddress,footinbib,citeautoscript]{revtex4-1}
\usepackage{graphicx} %
\usepackage{amsmath}
\usepackage{mathtools} %
\usepackage{bm}
\usepackage{amssymb} %
\usepackage{booktabs}
\usepackage{multirow}
\usepackage{mathrsfs}

\usepackage{siunitx} %
\usepackage{hyperref}
\hypersetup{colorlinks=true,linkcolor=blue,citecolor=blue,urlcolor=blue}

\usepackage{braket} %
\newcommand{\ketbra}[2]{\ket{#1}\!\bra{#2}}

\newcommand{\eu}{\mathrm{e}^}

\newcommand{\rmd}{\mathrm{d}}
\newcommand{\thalf}{{\ensuremath{\tfrac{1}{2}}}} %
\newcommand{\half}{{\ensuremath{\frac{1}{2}}}}

\newcommand{\op}[1]{\ensuremath{\hat{#1}}}
\renewcommand{\mathbf}[1]{\bm{#1}}
\providecommand{\mat}[1]{\mathbf{#1}}
\newcommand{\del}{\nabla}
\DeclareMathOperator{\tr}{tr}

\newcommand{\T}{\mathsf{T}}

\newcommand{\grad}{\mathbf{\del}}

\newcommand{\der}[3][]{\frac{\rmd^{#1}{#2}}{\rmd{#3}^{#1}}}
\newcommand{\pder}[3][]{\frac{\partial^{#1}{#2}}{\partial{#3}^{#1}}}

\newcommand{\eqn}[1]{Eq.\,(\ref{#1})}
\newcommand{\eqs}[1]{Eqs.\,(\ref{#1})}
\newcommand{\fig}[1]{Fig.\,\ref{fig:#1}}

\newcommand{\tref}[1]{Table \ref{tab:#1}\@}
\newcommand{\secref}[1]{Sec.\,\ref{sec:#1}\@}
\newcommand{\Ref}[1]{Ref.~\onlinecite{#1}}

\newcommand{\gam}[1]{\mathbf{\gamma_#1}}
\newcommand{\cmin}[1]{#1^\star} %

\graphicspath{{data+plots/}{presentation_plots/}}

\DeclareSIUnit{\mol}{mol}
\DeclareSIUnit{\cal}{cal}
\DeclareMathOperator{\csch}{csch}

\DeclareMathOperator{\arsinh}{arsinh}
\newcommand{\hartree}{E_\text{h}}

\newcommand{\tauinst}{\tau_\text{inst}}
\newcommand{\xinst}{x_\text{inst}}
\newcommand{\qinst}{q_\text{inst}}
\newcommand{\dotqinst}{\dot{q}_\text{inst}}
\newcommand{\Sinst}{S_\text{inst}}

\begin{document}

\title{Nonadiabatic quantum transition-state theory in the golden-rule limit. I.\ Theory and application to model systems}

\author{Manish J. Thapa}
\author{Wei Fang}
\author{Jeremy O. Richardson}
\email{jeremy.richardson@phys.chem.ethz.ch}
\affiliation{Laboratory of Physical Chemistry, ETH Z{\"u}rich, 8093 Z{\"u}rich, Switzerland}

\date{\today}

\begin{abstract}
\noindent
We propose a new quantum transition-state theory for calculating Fermi's golden-rule rates in complex multidimensional systems.
This method is able to account for the nuclear quantum effects of delocalization, zero-point energy and tunnelling
in an electron-transfer reaction.
It is related to instanton theory but
can be computed by path-integral sampling
and is thus applicable to treat molecular reactions in solution.
A constraint functional based on energy conservation is introduced which ensures that
the dominant paths contributing to the reaction rate are sampled.
We prove that the theory gives exact results for a system of crossed linear potentials
and also the correct classical limit for any system.
In numerical tests,
the new method is also seen to be accurate for anharmonic systems,
and even gives good predictions for rates in the Marcus inverted regime.
\end{abstract}

\maketitle

\section{Introduction}

One of the most successful approaches
for deriving new quantum transition-state theories \cite{Gillan,Voth+Chandler+Miller,Mills1997QTST,RPInst,Hele2013QTST,*Althorpe2013QTST,*Hele2013unique,Miller2003QI,Pollak1998QTST}
is by connection to semiclassical instanton rate theory \cite{Miller1975semiclassical,Benderskii,Perspective}.
This theory
can be rigorously derived as an asymptotic approximation to the exact rate constant
\cite{AdiabaticGreens,InstReview},
and the ring-polymer representation of the instanton can be employed to study polyatomic reactions in the gas phase
\cite{Andersson2009Hmethane,RPInst,Rommel2011locating}.
This makes semiclassical instanton theory
a remarkably efficient and accurate method for
studying quantum tunnelling in molecules and clusters.
\cite{Milnikov2001,Kaestner2014review,HCH4,water,hexamerprism,formic,GPR,porphycene,broadtop,Asgeirsson2018instanton,Cvitas2018instanton}

For the atomistic simulation of reactions in solution,
the semiclassical instanton method is no longer applicable,
but an alternative method is provided by 
ring-polymer molecular dynamics (RPMD).
\cite{RPMDcorrelation,RPMDrate,RPMDrefinedRate,Habershon2013RPMDreview}
The excellent accuracy achieved by RPMD rate theory
can be explained by its
connection to instanton theory, 
as the instanton can be shown to be equivalent to the dominant ring-polymer configuration sampled by the RPMD scheme.\cite{RPInst}
RPMD can of course also be used to simulate gas-phase reactions
\cite{RPMDgasPhase,*RPMDgasPhaseErratum,Suleimanov2016rate}
where it has been shown that it gives a similar level of accuracy to that of the instanton \cite{DMuH}.

In this paper, %
we shall study reactions in the nonadiabatic limit,
for which a typical case is 
electron transfer 
accompanied by a reorganization of the solvent environment \cite{ChandlerET}.
A simulation requires the consideration of at least two electronic states,
and hence one cannot use the Born-Oppenheimer (BO) approximation to study these processes.
Quantifying the rate of such reactions is challenging, but instrumental in understanding redox processes in chemical and biological systems. \cite{Blumberger2015ET}
Nuclear quantum effects such as tunnelling can play a significant role in these reactions
\cite{Bader1990golden,HammesSchiffer2015PCET,Pollak2012tunnel},
as has been experimentally confirmed by large room-temperature kinetic isotope effects
\cite{Hu2014enzyme}.

The celebrated Marcus theory
\cite{Marcus1964review,Marcus1985review,Marcus1993review}
provides an approximate solution to the rate of an electron-transfer reaction
by assuming
that the nuclei can be described by classical statistical mechanics
and that the resulting free-energy curves are harmonic.
Although extensions have been proposed to extend Marcus theory beyond 
the classical approximation,
many of them rely on an underlying assumption that the system can be described by a spin-boson model.
\cite{Siders1981quantum,*Siders1981inverted}
Hence there is a need for a full-dimensional atomistic simulation approach
which is applicable to study electron-transfer processes in liquid systems including nuclear quantum effects.
We shall derive a new quantum transition-state theory to treat such processes,
using a connection to semiclassical instanton theory to guide our derivation.

In this paper, we shall consider the rate of population transfer from the reactant electronic state, $\ket{0}$, to the product electronic state, $\ket{1}$,
at a temperature defined by $\beta=(k_\text{B}T)^{-1}$.
The total Hamiltonian of the system is given by
\begin{align}
	\op{H} = \op{H}_0 \ketbra{0}{0} + \op{H}_1 \ketbra{1}{1} + \Delta \big( \ketbra{0}{1} + \ketbra{1}{0} \big).
\end{align}
Each diabatic state, $\ket{n}$, has an associated nuclear Hamiltonian,
$\op{H}_n = \op{p}^2/2m + V_n(\op{x})$, %
which for the purpose of simplifying the presentation
we shall first consider to be one-dimensional
and extend the approach to multidimensional systems in \secref{multidim}.
We shall assume that the golden-rule limit applies,
in which the coupling, $\Delta$, between these electronic states is weak. %
Fermi's golden-rule rate constant is defined as \cite{fermiGR} 
\begin{align}
	\label{k}
	k = \frac{2\pi\Delta^2}{\hbar} \frac{Z^\ddag}{Z_0},
\end{align}
where $Z_0=\tr\big[\eu{-\beta\op{H}_0}\big]$ is the reactant partition function.
The transition-state partition function is defined as \cite{Miller1993QTST}
\begin{align}
	Z^\ddag &= \int \eu{-\beta E} \tr\left[\delta(\op{H}_0-E) \, \delta(\op{H}_1-E)\right] \rmd E
	\label{exactE}
	\\
	&= \int \tr\left[\eu{-\op{H}_0\tau_0/\hbar} \, \delta(\op{H}_0-E) \, \eu{-\op{H}_1\tau_1/\hbar} \, \delta(\op{H}_1-E)\right] \rmd E \,,
	\label{exacttau}
\end{align}
where $\tau_0\equiv\tau$ and $\tau_1\equiv\beta\hbar-\tau$
are imaginary times
corresponding to a particular splitting of the Boltzmann operator between the reactant and product states.
We shall employ these definitions for $\tau_n$ in terms of $\tau$ and $\beta$ throughout this paper.
Note that because this expression forces the reactant and product energies to be equal,
the exact golden-rule rate is independent of the choice of $\tau$
\cite{Miller1983rate}.

In previous work, we have rigorously derived a semiclassical instanton expression for the rate constant of this reaction \cite{GoldenGreens,InstReview}.
This approach includes the important quantum effects of tunnelling, delocalization and zero-point energy within an $\hbar\to0$ asymptotic approximation.
We have shown how the instanton can be written as a discretized ring polymer \cite{GoldenRPI}, which makes numerical evaluation of the rate computationally efficient,
and have successfully tested this method on an asymmetric system which generalizes the spin-boson model such that the free-energy curves become anharmonic \cite{AsymSysBath}.

The instanton is formed of two imaginary-time classical trajectories,
one on each state, joined together to form a closed ring. %
The length of time the trajectory travels on the reactant surface is $\tau_0$ 
whereas its time on the product surface is $\tau_1$.
The total time of the instanton path is thus $\beta\hbar$,
as in other semiclassical theories which involve the Boltzmann distribution
\cite{Miller1971density,Miller1975semiclassical}.
In the instanton approach, the time $\tau=\tauinst$ is chosen such that %
the energies of the two trajectories match.
The instanton must therefore always be located in the vicinity of the crossing seam, where $V_0(x)=V_1(x)$,
and in fact the
two trajectories join together smoothly 
on the crossing seam to form a periodic orbit.
Instanton rate theory is exact for a system of crossed linear potentials, as defined in Appendix \ref{sec:linear}\@. 
In the classical limit, it is well behaved 
and gives a steepest-descent approximation to the true classical rate.
\cite{GoldenGreens}

The main disadvantage of semiclassical instanton theories is that they cannot be directly applied to systems with explicit solvents.
This is because there would be such a large number of instanton orbits,
each corresponding to a different configuration of the solvent, that they could not all be located.
Also, if the instantons are close to each other in configuration space,
the individual steepest-descent approximations will overlap with each other and give a poor estimate of the configuration integral \cite{BenderBook}.
Taking inspiration from quantum simulations of liquids, \cite{Chandler+Wolynes1981,RPMDdiffusion,*Miller2005water,Ceriotti2016water,Markland2018review}
one sees that a better method would sample over relevant path-integral configurations in order to obtain the rate.

Wolynes has developed a theory to predict the golden-rule rate from a path-integral Monte Carlo simulation
\cite{Wolynes1987nonadiabatic},
which has been employed with atomistic simulations
in a number of applications  
\cite{Zheng1989ET,*Zheng1991ET,Bader1990golden}.
The theory was derived %
as a steepest-descent approximation to the flux-flux correlation function with respect to the time variable only.
This is very similar to the derivation of the so-called ``quantum instanton'' theory in the adiabatic limit described in \Ref{Vanicek2005QI}.
The same theory can also be obtained by analytic continuation of the imaginary free-energy \cite{Cao1997nonadiabatic,*Cao1998erratum}.
The resulting method can be computed in terms of an integral %
over imaginary-time paths, $x(t)$, with the cyclic boundary conditions, $x(0)=x(\beta\hbar)$.
The paths are split over the two potential-energy surfaces according to the variable $\tau$
and have action $S_\tau[x(t)]$, which is defined later in \eqn{S}.
The value of $\tau$ is chosen to minimize the integral. %
However, Wolynes theory is not directly connected to semiclassical instanton theory,
and as well as sampling configurations near to the crossing seam, it also includes configurations far away.
Like the adiabatic quantum instanton approach \cite{Miller2003QI},
it does not tend to the correct classical limit in general anharmonic systems \cite{GoldenRPI}.
We are thus interested in deriving a new path-integral sampling scheme that demonstrates a stronger connection to semiclassical instanton theory.

We would like to obtain a new approach which samples over paths
subject to a constraint such that the integral is
dominated by the instanton-like configurations
near the crossing seam.
Like other quantum transition-state theories %
\cite{Gillan,Voth+Chandler+Miller,Mills1997QTST,RPInst,Hele2013QTST},
it should not perform real-time propagation
but be based only on a constrained imaginary-time path-integral simulation.
In adiabatic quantum transition-state theories, the constraint is typically the centroid of the path integral,\cite{Gillan,Voth+Chandler+Miller} or generalizations of this which include higher normal modes \cite{Mills1997QTST,RPInst,Hele2013QTST}.
In \Ref{RPInst}, it was shown that the optimal dividing surface passes through the instanton
and in this way the semiclassical instanton rate can be related to quantum transition-state theory and hence to RPMD.

In order to obtain such a method, we shall need to introduce a constraint functional which
is obeyed by instanton configurations.
Following \Ref{nonoscillatory}, we suggest using a constraint
which forces the energy of the reactant and product states to be equal.
We therefore propose a new theory, called golden-rule quantum transition-state theory (GR-QTST),
in which the rate constant is defined by \eqn{k} with the ansatz
\begin{align}
	\label{approx}
	Z^\ddag \approx \beta \, \eu{-\phi(\tau^{*})/\hbar},
\end{align}
where, written in path-integral notation \cite{Feynman}, the effective action, $\phi(\tau)$, is defined as
\begin{align}
	\label{phi}
	\eu{-\phi(\tau)/\hbar} &= \oint \eu{-S_\tau[x(t)]/\hbar} \, \delta(\sigma_\tau[x(t)]) \, \mathcal{D} x(t).
\end{align}
The dimensionless constraint functional, $\sigma_\tau[x(t)]$, will be defined below
such that it forces the energy of the reactant and product to match
for every path in the ensemble.
Our derivation will rely on the idea that the method should predominantly sample the instanton trajectory and fluctuations nearby.
Therefore only paths in the region around the crossing seam,
including the instanton configuration itself,
should be located on the constraint hypersurface, $\sigma_\tau[x(t)]=0$.

In this paper, we shall obtain a definition for an appropriate constraint functional which has these required properties.
Rather than variationally optimizing a constraint based on centroids and ring-polymer normal modes,
we shall use the physically suggestive constraint that the energy of the reactants and products must be equal. 
We shall show how the resulting method tends to the correct classical limit for high temperatures or heavy masses.
A ring-polymer discretization, which makes the method applicable to complex multidimensional systems, will also be presented.
We shall investigate a number of model systems for which exact results are available for comparison
and will apply the method to atomistic simulations in future work.

\section{Theory}
\label{sec:theory}

In this section, we define the action functional, $S_\tau[x(t)]$, and derive a constraint functional, $\sigma_\tau[x(t)]$, with the properties required for the new GR-QTST method.
In order to make the argument as clear as possible, we first treat a one-dimensional system.
The advantage of using path-integral approaches is that they easily generalize to multidimensional systems
and the necessary extra steps are outlined in \secref{multidim}.

We shall employ the Lagrangian formalism of classical mechanics
and its extension to quantum mechanics provided by Feynman's path-integral theory \cite{Feynman}.
The Lagrangian for imaginary-time dynamics in the electronic state $\ket{n}$, corresponding to the Hamiltonian $\op{H}_n$, is
\begin{align}
	L_n \equiv L_n(x,\dot{x}) = \thalf m \dot{x}^2 + V_n(x).
\end{align}
Note that the sign of the second term is opposite from that of the usual expression
as it is known that imaginary-time dynamics is equivalent to classical dynamics with an inverted potential \cite{Miller1971density}.

The most important functional is the Euclidean action, which is defined as
\begin{align}
	S_n[x(t)] &= \int_0^{\tau_n} L_n\big(x(t),\dot{x}(t)\big) \, \rmd t.
\end{align}
A path for which $S_n[x(t)]$ is stationary
is called a trajectory and is denoted $\tilde{x}(t)$. \cite{GutzwillerBook}

Two open-ended paths are combined into a ring, $x(t)$,
such that $x(0)=x(\beta\hbar)$,
where imaginary time, $t$, runs from $0$ to $\tau$ on the reactant state
and from $\tau$ to $\beta\hbar$ on the product.
The total action is the sum of the individual actions of each of these paths,
\begin{align}
	\label{S}
	S_\tau[x(t)] &= S_0[x(t)] + S_1[x(\tau+t)],
\end{align}
where the time integral is from $0$ to $\tau$ on the reactant and from $\tau$ to $\beta\hbar$ on the product.
The coordinates where the path hops from one state to another are
\begin{align}
	x' &\equiv x(0),
	&
	x'' &\equiv x(\tau).
\end{align}
Solving a set of equations,
$\pder{}{x'}S_{\tau}[\tilde{x}(t)]=0$, $\pder{}{x''}S_{\tau}[\tilde{x}(t)]=0$ and $\pder{}{\tau}S_{\tau}[\tilde{x}(t)]=0$
yields a special trajectory, with $\tilde{x}(t)=\xinst(t)$ and $\tau=\tauinst$. This classical trajectory is called the
instanton and it hops at $x_\text{inst}'=x_\text{inst}''=x^\ddag$, where $x^\ddag$ is defined as the crossing point of the diabatic surfaces, such that $V_0(x^\ddag)=V_1(x^\ddag)$.
That the derivative of the action with respect to $\tau$ is zero
ensures that the two trajectories of the instanton have the same energy. \cite{GoldenGreens}
Our formulation of GR-QTST is motivated by a similar energy-matching constraint.

As will become apparent, it will be necessary to rewrite the Lagrangian in terms of a generalized coordinate, $q\equiv q(x)$.
The Euler-Lagrange equation which is satisfied by trajectories, $\tilde{q}(t)\equiv q(\tilde{x}(t))$, is
\begin{align}
	\label{EL}
	\pder{}{t} \left( \pder{L_n}{\dot{q}} \right) &= \pder{L_n}{q}
	& \text{for } q=\tilde{q} \text{ and } \dot{q}=\dot{\tilde{q}},
\end{align}
where the conjugate momentum of the new coordinate is
\begin{align}
	\pder{L_n}{\dot{q}} &= m \dot{x} \pder{\dot{x}}{\dot{q}} = m \dot{x} \pder{x}{q} 
\end{align}
and the right-hand side is
\begin{align}
	\pder{L_n}{q} &= m \dot{x} \der{}{t} \left(\pder{x}{q}\right) + \pder{V_n}{q} 
	\nonumber \\
	&= - m \dot{x} \left(\pder{q}{x}\right)^{-1} \pder[2]{q}{x} \left(\pder{q}{x}\right)^{-1} \dot{x} + \pder{V_n}{x} \left(\pder{q}{x}\right)^{-1}.
	\label{dLdq}
\end{align}

Following standard manipulations discussed in many classical mechanics text books,
it can be shown that the
instantaneous energy, here defined as $E_n(q,\dot{q})=-\pder{L_n}{\dot{q}}\dot{q}+L_n$, 
is conserved along a trajectory, $\tilde{q}(t)$,
as its time-derivative is zero if the Euler-Lagrange equation is obeyed.
There is, however, no unique definition for a functional reporting on the energy of a non-classical path, $q(t)$.
One possible measure is 
the average of the instantaneous energy along the path:
\begin{subequations}
\label{Ebar}
\begin{align}
	\bar{E}_n^\text{th}[x(t)]
	&= \frac{1}{\tau_n} \int_0^{\tau_n} E_n(q,\dot{q}) \, \rmd t
	\\
	&= \frac{1}{\tau_n} \int_0^{\tau_n} \left[ -\half \pder{L_n}{\dot{q}} \dot{q} + V_n(x) \right] \rmd t.
	\label{Eth}
\end{align}
\end{subequations}
This is known as the thermodynamic estimator as it can also be derived from the derivative with respect to $\beta$ of the partition function.
There are however other functionals which can be designed to report on the energy of a path.
The only requirement is that they return the correct energy for trajectories,
but may give different results for a non-classical path. 

At first, we attempted to use the thermodynamic estimator, \eqn{Ebar}, to define the constraint functional
\begin{align}
	\sigma_\tau^{\text{th}}[x(t)] = \beta \left( \bar{E}_0^\text{th}[x(t)] - \bar{E}_1^\text{th}[x(\tau+t)] \right),
\end{align}
but found that although the instanton is located on the constraint hypersurface,
it also allows non-instanton paths to dominate the sampling,
as shown in \fig{FreeEnergy}.
The thermodynamic energy contains a minus sign in front of the kinetic-energy term
and thus this constraint may allow for sampling of highly-stretched paths which are unphysical.
To avoid this problem, it is necessary to remove all terms involving $\dot{x}$.

Thus we shall use classical-mechanical considerations to
obtain a different energy functional in such a way
that it does not change the result for the instanton trajectories,
but which is not defined in terms of $\dot{x}$.
This follows a similar derivation to the virial theorem in classical statistical mechanics
and a similar strategy has been applied previously to path-integrals to get virial energy estimator.
\cite{Herman1982PIMC,Yamamoto2005virial,Karandashev2015QI}

First we integrate \eqn{Eth} by parts to get
\begin{align}
	\bar{E}_n^\text{th}[x(t)]
	&= -\frac{1}{2\tau_n}\left. \pder{L_n}{\dot{q}}q \right|_0^{\tau_n}  
	\nonumber\\&\quad+\frac{1}{\tau_n} \int_0^{\tau_n} \left[ \half \der{}{t}\left(\pder{L_n}{\dot{q}}\right) q + V_n(x) \right] \rmd t.
\end{align}
The first term in the integrand will be modified by applying
the Euler-Lagrange equation, \eqn{EL}, even though it is not obeyed by a non-classical path.
We shall also choose to define $q(x)$ such that for the instanton trajectory
$q'_\text{inst}\equiv q(x_\text{inst}')=0$ and $q''_\text{inst}\equiv q(x_\text{inst}'')=0$,
and thus the boundary terms can be neglected.
This gives the definition of the virial energy functional,
\begin{align}
	\bar{E}_n^\text{v}[x(t)]
	&= \frac{1}{\tau_n} \int_0^{\tau_n} \left[ \half \pder{L_n}{q} q + V_n(x) \right] \rmd t.
\end{align}
The two energy estimators are
equal only for an instanton trajectory:
i.e.\ $\bar{E}_n^\text{th}[\xinst(t)] = \bar{E}_n^\text{v}[\xinst(t)]$. %
Their value is also equal to the instantaneous classical energy $E_n(\qinst,\dot{q}_\text{inst})$ at all points along the instanton trajectory.

One can now see that it was necessary to change to the generalized coordinate system
so that we could ignore the boundary term.
In order to choose the form of $q(x)$ correctly, we take into consideration all knowledge that we have about the instanton in general.
We have already discussed that the energies of the two trajectories are equal, $E_0=E_1$, and are using this fact to define the constraint.
We also know that the instanton hops between electronic states at the crossing point where $V_0(x)=V_1(x)$,
and this information can be used to obtain a definition for $q(x)$.

One choice for the generalized coordinate might be $q(x) = V_-(x) \equiv V_0(x) - V_1(x)$.
This obviously goes to zero at the ends of the instanton as required.
The definition also has a physical interpretation as this is often used as the reaction coordinate in classical simulations of electron-transfer reactions
\cite{ChandlerET,Hwang1987ET,*King1990ET,Kuharski1988Fe3e}.
However, in general, because $\pder[2]{q}{x} \ne 0$
it
will give a complicated form for $\pder{L_n}{q}$ (see \eqn{dLdq}) which involves the $\dot{x}$ terms which we are trying to remove.
A better choice is its linearized form:
\begin{align}
	\label{q}
	q(x) = V_-(x_+) + \del V_-(x_+) (x - x_+),
\end{align}
which is a first-order Taylor expansion around $x_+\equiv \thalf(x'+x'')$.
This is also a suitable definition for the generalized coordinate
as it clearly goes to 0 at the hopping points of the instanton as required.
Note that this approach does not require knowledge of the location of the crossing seam
which would not in general be known for a complex multidimensional system.

This generalized coordinate transform
gives a simple form for $\pder{L_n}{q} = \del V_n(x) \big(\del V_-(x_+)\big)^{-1}$
and hence
\begin{align}
	\bar{E}_n^\text{v}[x(t)] = \frac{1}{\tau_n} \int_0^{\tau_n} \left[
		\half \del V_n(x) \big( x - s(x_+) \big) + V_n(x) \right] \rmd t ,
\end{align}
where $s(x_+) = x_+ - \frac{V_-(x_+)}{\del V_-(x_+)}$.
For systems where $V_-(x)$ is linear, which includes the spin-boson model,
$s(x_+)$ will be defined at exactly the crossing point, $x^\ddag$, for all paths.

We choose to define the constraint functional for GR-QTST in terms of the virial energy as
\begin{align}
	\label{sigma}
	\sigma_\tau[x(t)] &= \tfrac{2}{3} \beta \big( \bar{E}_0^\text{v}[x(t)] - \bar{E}_1^\text{v}[x(\tau+t)] \big).
\end{align}
The factor of $\beta$ is included in the definition so as to make the functional dimensionless
and the factor of $\frac{2}{3}$ ensures that the method tends to certain important limits as shown in Appendices \ref{sec:classical}\@ and \ref{sec:linear}\@.

The GR-QTST proposed in this paper gives the correct rate for all systems in the classical limit, as shown in Appendix \ref{sec:classical}\@.
It gives the exact quantum rate in the case of a system of crossed linear potentials.
This is proved in Appendix \ref{sec:linear}\@.
This suggests that within the ansatz of our new method, we have chosen the correct functional form for $\sigma_\tau[x(t)]$.
Had we, for instance, allowed the boundary terms to remain or not performed the virial transformation,
the proof would no longer hold.
Note that our choice of coordinate transform is however not unique in general.
There will be other definitions of $q(x)$ with the correct properties
which in the case of a linear model or classical limit reduce to give the correct result.
In future work, we shall test other forms of the generalized coordinate to explore possible improvements to the method.

\section{Ring-polymer representation}
\label{sec:ringpolymer}

In order to apply the GR-QTST method to
complex systems for which analytical results are not available, one discretizes paths into $N$ ring-polymer beads,
$\mathbf{x}=\{x_1,\dots,x_N\}$. \cite{Chandler+Wolynes1981}
Between neighbouring beads, there are $N_0$ intervals describing dynamics of the reactant system
and $N_1\equiv N-N_0$ of the product system.
The functionals become functions of the positions of the ring-polymer beads,
which we obtain by using finite differences for time derivatives and the trapezoidal rule on the integrals over time.
In this representation, the total action, equivalent to \eqn{S}, is given by
\begin{align}
	\label{Srp}
	S_\tau(\mathbf{x})
	&= \sum_{i=1}^{N_0} \frac{m N_0}{2\tau_0} |x_i - x_{i-1}|^2 + \frac{\tau_0}{2N_0} \big( V_0(x_i) + V_0(x_{i-1}) \big)
	\nonumber\\&\quad + \sum_{i=N_0+1}^{N} \frac{m N_1}{2\tau_1} |x_i - x_{i-1}|^2 
	\nonumber\\&\quad + \frac{\tau_1}{2N_1} \big( V_1(x_i) + V_1(x_{i-1}) \big)
\end{align}
with cyclic indices such that $x_0\equiv x_N$.

With this definition, there are 
two simple discretization schemes.
The first is defined to keep the $N_0\!:\!N_1$ ratio fixed as $\tau$ is varied,
which leads to unequal spring force constants around the ring polymer. 
This is the scheme used in our ring-polymer instanton approaches. \cite{GoldenRPI,AsymSysBath}
The second is to vary the ratio $N_0\!:\!N_1$ with $\tau$ so as to preserve the equality $\tau_0/N_0\equiv\tau_1/N_1$.
In the limit of $N\rightarrow\infty$, the same results will be obtained by either method
although the convergence with respect to $N$ appears to be slightly better for the first approach, especially when $\tau$ approaches its extremes of 0 or $\beta\hbar$.
However, the second approach,
which is similar to the discretization scheme proposed by Wolynes \cite{Wolynes1987nonadiabatic},
may be simpler to implement in a computationally efficient atomistic simulation and is the approach employed in this paper.

The ring-polymer version of the constraint function, \eqn{sigma}, is
\begin{align}
	\label{sigrp}
	\sigma_\tau(\mathbf{x}) = \tfrac{2}{3}\beta \left( \bar{E}_0^\text{v}(x_0,\dots,x_{N_0}) - \bar{E}_1^\text{v}(x_{N_0},\dots,x_N) \right)
\end{align}
with the virial energy on each state defined by
\begin{align}
	\bar{E}_n^\text{v}(x_0,\dots,x_{N_n}) &= \frac{1}{2N_n} \sum_{i=1}^{N_n} 
		\bigg[
		V_n(x_i) + V_n(x_{i-1})
		\nonumber\\&\quad
		+ \half \del V_n(x_i) \big(x_i - s(x_+)\big)
		\nonumber\\&\quad
		+ \half \del V_n(x_{i-1}) \big(x_{i-1} - s(x_+)\big)
		\bigg],
	\label{Ev}
\end{align}
where $x_+ \equiv \thalf (x_0 + x_{N_0})$.
There may of course be other finite-difference formulae which have better convergence with $N\rightarrow\infty$,
but we found that this one worked well for a reasonably small value of $N$.

The path integral in \eqn{phi} becomes the following constrained configuration integral over the ring polymer:
\begin{align}
	\label{phiRP}
	\eu{-\phi(\tau)/\hbar} &= \Lambda^{-N} \int \eu{-S_\tau(\mathbf{x})/\hbar} \, \delta(\sigma_\tau(\mathbf{x})) \, \rmd \mathbf{x},
\end{align}
with the normalization constants
$\Lambda=\sqrt{2\pi\beta\hbar^2/mN}$. \cite{Feynman}
The rate constant of the GR-QTST method is then defined by \eqn{k} with the approximation of \eqn{approx} and the definitions \eqs{sigrp} and (\ref{phiRP}).
The reactant partition function is defined in the usual way as
\begin{align}
	Z_0 = \Lambda^{-N} \int \eu{-S_{\beta\hbar}(\mathbf{x})/\hbar} \, \rmd \mathbf{x},
\end{align}
with $N_0=N$ and $N_1=0$.

The GR-QTST method is defined using the virial energy estimator.
However, for comparison, we note that the thermodynamic estimator in ring-polymer form is
\begin{align}
	\bar{E}^{\text{th}}_n(x_0,\dots,x_{N_n}) &= \frac{1}{2N_n}\sum_{i=1}^{N_n} \bigg[ V_n(x_i)+V_n(x_{i-1})
	\nonumber\\&\quad-\frac{mN_n^2}{\tau_n^2} |x_i-x_{i-1}|^2 \bigg].
\end{align}
This is used to define the thermodynamic constraint, %
which however should not be used in the GR-QTST ansatz as we shall show.

Wolynes theory \cite{Wolynes1987nonadiabatic} is based on an \emph{unconstrained} path-integral simulation
and defines the rate constant by \eqn{k} with the ansatz
\begin{subequations}
\label{Wolynes}
\begin{align}
	Z^\ddag &\approx \left[-2\pi\hbar \, \pder[2]{\phi_\text{u}}{\tau} \right]^{-\half}_{\tau=\tau^*} \, \eu{-\phi_\text{u}(\tau^*)/\hbar}
	\\
	\eu{-\phi_\text{u}(\tau)/\hbar} &= \Lambda^{-N} \int \eu{-S_\tau(\mathbf{x})/\hbar} \, \rmd \mathbf{x},
\end{align}
\end{subequations}
where $\tau^*$ is chosen to maximize the effective action, $\phi_\text{u}(\tau^*)$.

\begin{figure}[ht!]
	\includegraphics[width=3.37 in]{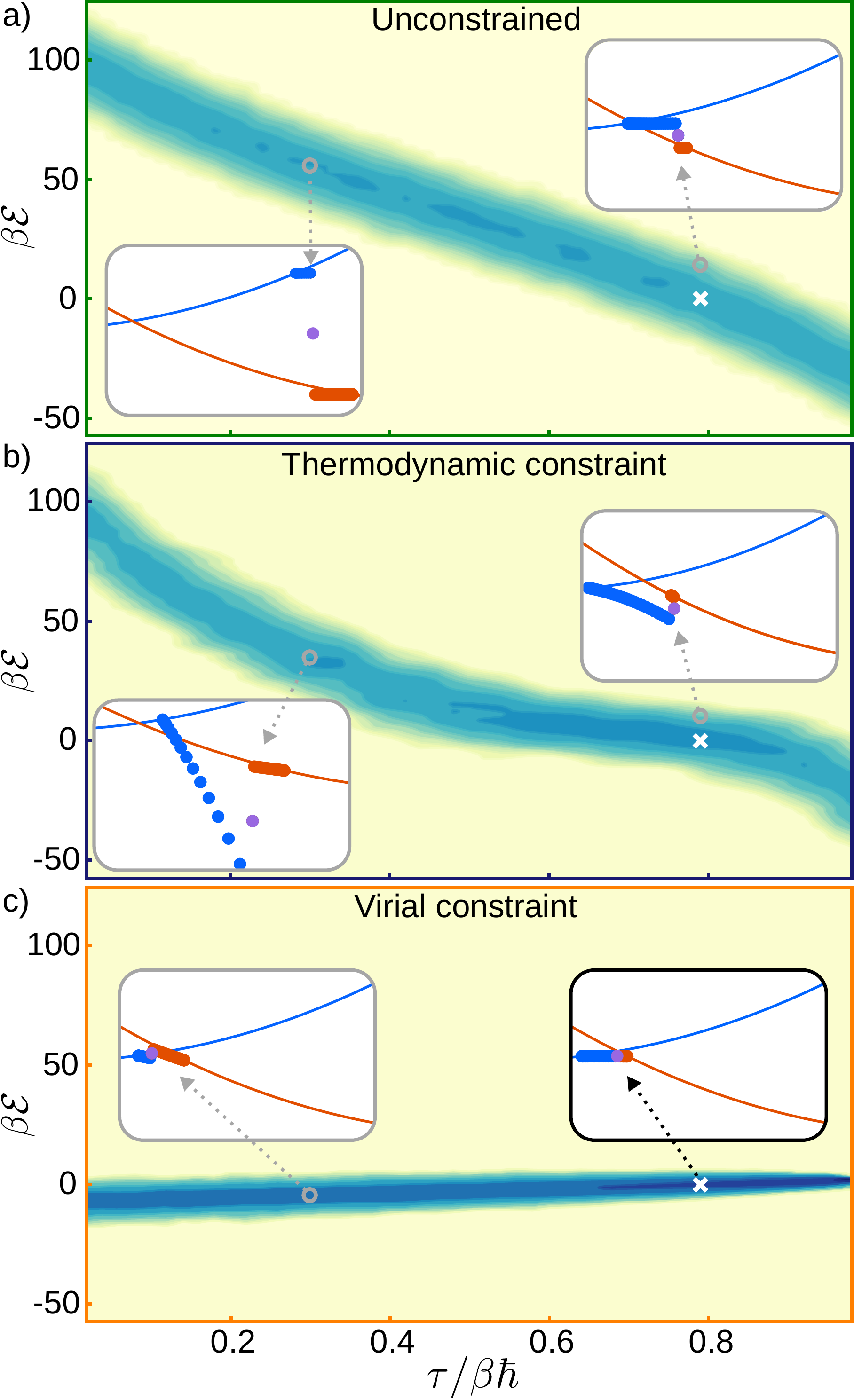}
	\caption{Density plots of the relative free energy, $F_{\tau}(\mathcal{E})=-\log p_{\tau}(\mathcal{E})$, over a range of $\tau$ values.
$p_{\tau}(\mathcal{E})$ is the probability density of 
finding the average hopping point, $x_+$, with the polarization $\mathcal{E}\equiv\thalf V_-(x_+)$
at a given $\tau$ obtained from path-integral sampling with
(a) no constraint, (b) the thermodynamic constraint, and (c) the virial constraint.
(Note that here the probability is normalized for each value of $\tau$ such that $\int p_{\tau}(\mathcal{E})\,\rmd \mathcal{E}=1$.)
This system is the one-dimensional 
harmonic model defined in \secref{sysbath}, with $\epsilon/\lambda = 0.5$ and no bath modes.
For this system, $\tauinst/\beta \hbar \approx 0.79$.
The regions which are rarely sampled are coloured yellow whereas the commonly sampled regions are blue.
The crosses signify the instanton configurations.
The insets show the path with the minimal action at the indicated values of $\tau$ and $\mathcal{E}$.
The x-axis of the insets is the nuclear configuration, $x$, and 
the y-axis shows the potential-energy surfaces and the instantaneous energy of the beads.
}
	\label{fig:FreeEnergy}
\end{figure}

A probability distribution of the sampled configurations is plotted in \fig{FreeEnergy}
for both the thermodynamic and virial constraints, as well as for an unconstrained version.
As shown in the insets, even for values of $\tau$ far from $\tauinst$, configurations obtained with the virial constraint resemble
instanton-like configurations localized close to the crossing seam, $x^\ddag$.
However configurations sampled using either the thermodynamic constraint or without constraint
can be found far away.
As expected, the unconstrained simulation samples configurations which do not require that
the energy of beads on the reactant and product surfaces match
and which are thus clearly unphysical descriptions of the reaction.
In each case, simulations carried out at $\tau=\tauinst$ do sample the instanton,
but the unconstrained simulation as well as simulations performed using the thermodynamic
constraint also sample a much wider set of configurations which are not localized at the crossing seam.
Therefore results using these two approaches, unlike when using the virial constraint, may be contaminated by unphysical configurations.

We next study the behaviour of the action with respect to $\tau$.
There exists a point, $\cmin{\mathbf{x}}$, for each value of $\tau$ between 0 and $\beta \hbar$ which minimizes the action, $S_\tau(\cmin{\mathbf{x}})$, while obeying the constraint. 
The constrained minimization is performed by introducing
a Lagrange multiplier, $\mu$, and solving
the following $(N+1)$ equations:%
\footnote{MINPACK's subroutine \textit{hybrd}
was used to perform the constraint minimizations 
through a \textit{scipy} wrapper
to obtain $\cmin{\mathbf{x}}$ for different values of $\tau$ in the range $[0, \beta \hbar]$.
The algorithm uses a forward-difference approximation to calculate the Jacobian.
The initial guess was taken from the previously computed solution at a different value of $\tau$ or from the ring-polymer instanton configuration.
}
\begin{subequations}
\begin{align}
	\label{lagrangemulti}
	\grad S_\tau(\cmin{\mathbf{x}}) - \mu \grad \sigma_\tau(\cmin{\mathbf{x}}) &= \mat{0}
	\\
	\sigma_\tau(\cmin{\mathbf{x}}) &= 0 \, ,
\end{align}
\end{subequations}
where $\grad$ denotes the derivative with respect to $\mathbf{x}$.
Note that for $\tau=\tauinst$,
the solution $\cmin{\mathbf{x}}$
represents the ring-polymer instanton configuration, $\xinst(t)$, but is otherwise a non-classical path.

\begin{figure}
	\includegraphics{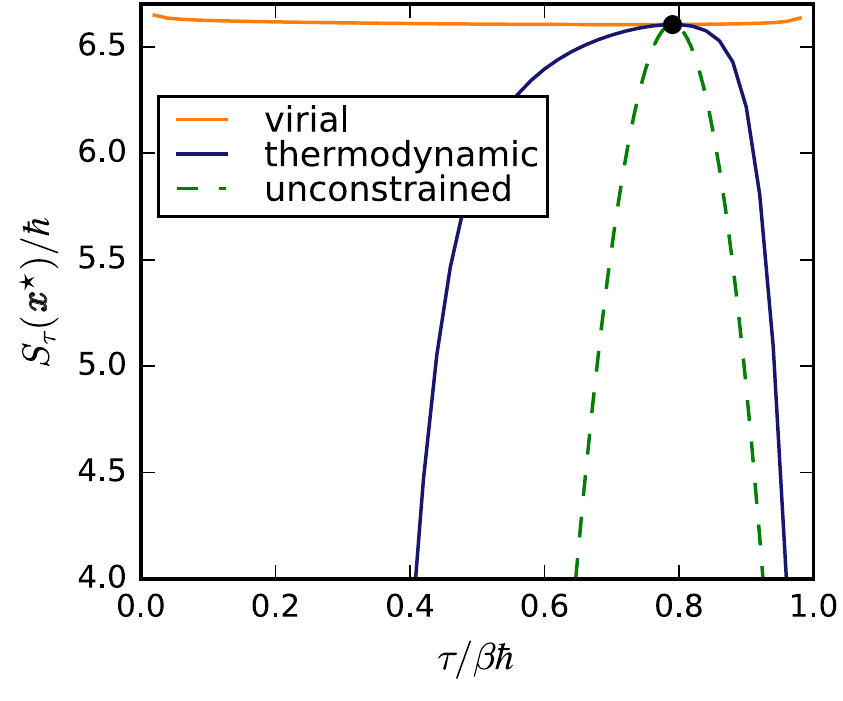}
	\caption{
	Minimized actions, $S_\tau(\cmin{\mathbf{x}})$, with and without constraints
	for various values of $\tau$ are compared
	for the one-dimensional harmonic system used in Figure \ref{fig:FreeEnergy}.
	The minimizations were performed with $N=50$.
	The black dot indicates the action corresponding to the ring-polymer instanton configuration computed using the same number of path-integral beads.
	}
	\label{fig:minimalS}
\end{figure}

As an example, we choose a one-dimensional biased system in which both potential-energy surfaces are harmonic.
In \fig{minimalS}, we plot the value of the action, \eqn{Srp},
after performing a constrained minimization for different values of $\tau$.
The function $S_\tau(\mathbf{x}^\star)$ is almost flat for this system,
and in fact,
as shown in Appendix \ref{sec:linear}\@,
it becomes perfectly flat in the limiting case of the crossed linear model.
When $\tau$ is equal to its value at the instanton, $\tauinst$,
the value of $S_\tau(\cmin{\mathbf{x}})$ is equal to the instanton action, $\Sinst$.

Had we minimized $S_\tau(\mathbf{x})$ without the constraint, %
this function would be strongly dependent on $\tau$, as shown by the dashed line.
Even with the thermodynamic energy constraint %
the action still varies over orders of magnitude depending on $\tau$.
Both Figures \ref{fig:FreeEnergy} and \ref{fig:minimalS} show clearly that
only by using the virial energy constraint can we say that there is a strong connection to the instanton result
for any value of $\tau$.

\section{Path-Integral Monte Carlo Implementation}

Here we employ a Monte Carlo importance sampling approach based on 
an extension of the ring-polymer transition-state sampling methods used in previous work \cite{RPMDrate,RPInst}.
It would also be possible to compute the GR-QTST integral 
using thermodynamic integration
and Metropolis Monte Carlo \cite{Ceperley1995PathIntegrals} or path-integral molecular dynamics schemes. \cite{Marx1996PIMD}
This will be necessary to extend our method to the atomistic description of liquids, and we shall discuss such applications in future studies.

\subsection{Importance sampling}
\label{sec:PIMCimplement}
After locating the constrained minimum, $\cmin{\mathbf{x}}$, for a given $\tau$ value, as described in \secref{ringpolymer},
we shall perform a coordinate transformation similar to a normal-mode transformation.
In order to do this, we start by expanding our total action as a Taylor series around this point:
\begin{align}
	S_\tau(\mathbf{x}) &= S_\tau(\cmin{\mathbf{x}}) + \grad S_\tau(\cmin{\mathbf{x}}) \cdot (\mathbf{x} - \cmin{\mathbf{x}})
	\nonumber\\&\quad+\thalf (\mathbf{x} - \cmin{\mathbf{x}}) \cdot \grad^2 S_\tau(\cmin{\mathbf{x}}) \cdot (\mathbf{x} - \cmin{\mathbf{x}}) + \cdots
	\label{Staylor}
\end{align}
and %
\begin{align}
	\sigma_\tau(\mathbf{x}) = \grad \sigma_\tau(\cmin{\mathbf{x}}) \cdot (\mathbf{x} - \cmin{\mathbf{x}}) + \cdots.
	\label{sigmataylor}
\end{align}
In what follows, we do not truncate the series, so no approximation is made.

The transformation shall be defined by the orthogonal matrix $\mat{U}$, 
such that the first column is a vector parallel to $\grad\sigma_\tau(\cmin{\mathbf{x}})$,
i.e.\ $U_{i1} = \del_i\sigma_\tau(\cmin{\mathbf{x}}) |\grad\sigma_\tau(\cmin{\mathbf{x}})|^{-1}$, and so all other column vectors are orthogonal to this and normalized.
\footnote{A simple way to obtain the $\mat{U}$ matrix is
to build a matrix with the required first column and set the other columns to be any linearly-independent vectors.
The resulting matrix can then be orthogonalized using a QR decomposition.}
We shall make use of the fact that the elements of the vector $\mat{U}^\T \grad\sigma_\tau(\cmin{\mathbf{x}})$ are all zero except the first,
i.e.\ $\sum_{i=1}^N U_{ij} \del_i\sigma_\tau(\cmin{\mathbf{x}})  = |\grad\sigma_\tau(\cmin{\mathbf{x}})| \delta_{1j}$. Then from \eqn{lagrangemulti}, we have $\mathbf{b} \equiv \mat{U}^\T \grad S_\tau(\cmin{\mathbf{x}}) = \mu \mat{U}^\T \grad\sigma_\tau(\cmin{\mathbf{x}})$
which implies that 
therefore $b_1 = \mu|\grad\sigma_\tau(\cmin{\mathbf{x}})|$ and
$b_k=0$ for $k=2,\dots,N$.

We have not yet defined the other column vectors of $\mat{U}$.
They will be chosen so as to diagonalize
a sub-matrix of $\mat{A} = \mat{U}^\T \grad^2 S_\tau(\cmin{\mathbf{x}}) \mat{U}$
such that $A_{k k'}=\delta_{k k'}a_k$ for $k=2,\dots,N$ and $k'=2,\dots,N$.
Because $\cmin{\mathbf{x}}$ is the constrained minimum, we know that $a_k>0$ for $k=2,\dots,N$.
The coordinate transformation is then defined as $\mathbf{y}\equiv\mat{U}^\T (\mathbf{x}-\cmin{\mathbf{x}})$,
which gives
\begin{subequations}
	\label{Sy}
\begin{align}
	S_\tau(\mathbf{x})
		&= S_\tau(\cmin{\mathbf{x}}) + \mathbf{b} \cdot \mathbf{y} + \thalf \mathbf{y} \cdot \mat{A} \cdot \mathbf{y} + \cdots
		\\
		&= S_\tau(\cmin{\mathbf{x}}) + b_1 y_1 + \thalf A_{11} y_1^2 + \sum_{k=2}^N A_{1k} y_1 y_k 
		\nonumber\\&\quad+\sum_{k=2}^N \thalf a_k y_k^2 + \cdots\,.
\end{align}
\end{subequations}

To obtain an efficient Monte Carlo importance sampling scheme
for the integral in \eqn{phiRP},
we change to the new coordinates and then multiply and divide by a set of Gaussian distributions to give
\begin{align}
\label{grqtst}
	\eu{-\phi(\tau)/\hbar}
	&= \Lambda^{-1} \prod_{k=2}^N \sqrt\frac{2\pi\hbar}{\Lambda^2a_k} \int \eu{-S_\tau(\mathbf{x})/\hbar}
\nonumber\\&\quad
	\times \eu{\frac{1}{2\hbar}\sum_{k=2}^Na_ky_k^2}
		\left|\pder{\sigma_\tau(\mathbf{x})}{y_1}\right|^{-1}
		P(\mathbf{y}) \, \rmd \mathbf{y},
\end{align}
where here the action $S_\tau(\mathbf{x})$ is the full form, \eqn{Srp}.
We have introduced the normalized distribution
\begin{align}
	P(\mathbf{y}) = \delta(y_1-\cmin{y}_1) \prod_{k=2}^N \frac{\eu{-a_ky_k^2/2\hbar}}{\sqrt{2\pi\hbar/a_k}}
\end{align}
and defined $\cmin{y_1}$ such that $\sigma_\tau\big(\mathbf{x}(\cmin{y_1},y_2,\dots,y_N)\big)=0$.
Finally, note that the term arising from the normalization of the delta function is given by
\begin{align}
	\pder{\sigma_\tau(\mathbf{x})}{y_1}
	&= \sum_{i=1}^N \del_i\sigma_\tau(\mathbf{x}) U_{i1}.
\end{align}
In this form,
the integral lends itself well to a Monte Carlo importance sampling evaluation with $P(\mathbf{y})$ defining the sampling distribution.
For each Monte Carlo step, $y_k$ is sampled from the normal distribution, $y_k\sim\mathcal{N}(0,\hbar/a_k)$, for $k=2,\dots,N$.
Then $y_1$ is set to $\cmin{y_1}$ which is found by a one-dimensional root search algorithm,
starting from a guess which solves the first-order equation $0=\sigma_\tau(\mathbf{x}_{\textnormal{MC}})+\grad\sigma_\tau(\mathbf{\mathbf{x}}_{\textnormal{MC}}) \cdot \mat{U} \cdot \mathbf{y}$, where  $\mathbf{x}_{\textnormal{MC}}=\mat{U} \cdot (0,y_2,\dots,y_N) + \mathbf{\cmin{x}}$.
For the simulations performed using the virial constraint, the guess obtained from solving this equation 
analytically was always very close to the correct root and thus there was no problem in obtaining $\cmin{y_1}$ numerically.

\begin{figure}
\includegraphics{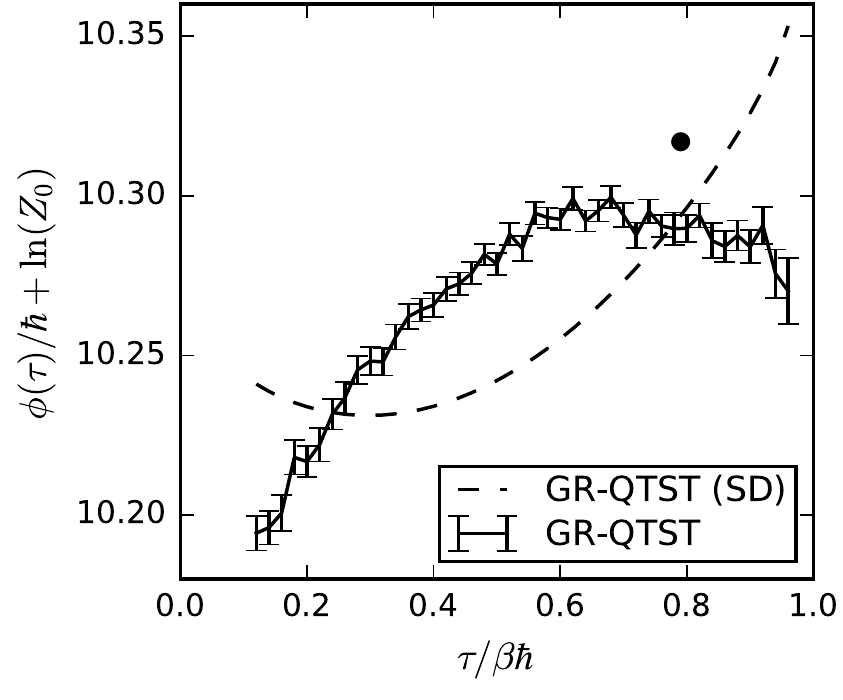}
	\caption{
	Plot showing the effective action for a one-dimensional harmonic system computed using GR-QTST, \eqn{grqtst},
	and its steepest-descent (SD) approximation, \eqn{sdapprox}.
	The results are scaled by the reactant partition function, $Z_0$,
	to ease comparison with the multidimensional system-bath model presented in \secref{sysbath}.
	The effective action, $\phi_{\textnormal{inst}}$, of the instanton method, \eqn{instphi}, is marked with a black dot
	at $\tau=\tauinst$.
	The system is the same as was used in Figures \ref{fig:FreeEnergy} and \ref{fig:minimalS}
	and simulations were performed with $N=50$.
	}
	\label{fig:phi1d}
\end{figure}

In the $N\rightarrow\infty$ limit
and within statistical error,
the Monte Carlo importance sampling scheme outlined above 
introduces no extra approximations into the GR-QTST method.
It is the approach which we shall use to test the method on model systems for which analytical results are not available.
However, 
one could also obtain a steepest-descent approximation to \eqn{grqtst}
by ignoring higher-order terms in \eqs{Staylor} and (\ref{sigmataylor}), which would be exact only for the linear system.
This gives $\cmin{y_1}=0$ and makes the exponent independent of $y_k$ for $k=2,\dots,N$\@.
Therefore
\begin{align}
\label{sdapprox}
	\eu{-\phi_\text{SD}(\tau)/\hbar}
	&= \Lambda^{-1} \prod_{k=2}^N \sqrt\frac{2\pi\hbar}{\Lambda^2a_k}
		\left|\grad\sigma_\tau(\cmin{\mathbf{x}})\right|^{-1}
		\eu{-S_\tau(\cmin{\mathbf{x}})/\hbar}.
\end{align}
Using \eqn{sdapprox} in \eqs{k} and (\ref{approx}) gives a steepest-descent approximation to the GR-QTST rate.

For comparison, 
under the ring-polymer instanton formalism,
the effective action, $\phi_\text{inst}$, is defined at $\tau=\tauinst$ as follows:\cite{GoldenGreens}
\begin{align}
\label{instphi}
	\eu{-\phi_\text{inst}/\hbar}
	&= \beta^{-1} \prod_{k=1}^N \sqrt\frac{2\pi\hbar}{\Lambda^2a_k}
		\left[-2 \pi\hbar\der[2]{\Sinst}{\tau}\right]^{-\half}_{\tau=\tauinst}
		\eu{-\Sinst/\hbar},
\end{align}
where $a_k$ are the eigenvalues of $\mathbf{A}$ and 
$\Sinst$ is the action corresponding to the instanton trajectory.
The instanton rate is given by \eqn{k} with the ansatz $Z^\ddag \approx \beta \, \eu{-\phi_\text{inst}/\hbar}$.
This formulation is clearly related to the steepest-descent version of GR-QTST, \eqn{sdapprox},
and has exactly the same exponential factor for $\tau=\tauinst$.
There is a minor difference in the prefactor, just as between a steepest-descent approximation to ring-polymer transition-state theory
and standard ring-polymer instanton rate theory. \cite{RPInst}

We are now in a position to make numerical calculations of GR-QTST for a general system.
We choose to illustrate the method using the example of the one-dimensional harmonic system,
for which results are shown in \fig{phi1d}.

The effective actions of GR-QTST and its steepest-descent approximation are very similar.
Although the curves of $\phi(\tau)$ and $\phi_\text{SD}(\tau)$ have 
a different shape due to subtle differences,
they are both 
seen to be almost independent of $\tau$ and within one decimal place give the same numerical value.
It is relatively unimportant which value of $\tau$ is used to define the GR-QTST rate
because $\phi(\tau)$ is approximately independent of this choice.
However, for reasons explained in Appendix \ref{sec:linearharmonic},
the ansatz of the new method is evaluated at $\tau^*$ which maximizes $\phi(\tau^*)$.

As expected, it is seen that the steepest-descent approximation, $\phi_\text{SD}(\tau)$,
is in excellent agreement with the instanton value, $\phi_\text{inst}$, for all values of $\tau$.
This is due to the similarity between \eqs{sdapprox} and (\ref{instphi}),
and the fact that the exponent, $S_\tau(\cmin{\mathbf{x}})$, is exactly equal to $\Sinst$ at $\tau=\tauinst$ and almost constant with respect to $\tau$ as shown in \fig{minimalS}.

Because of these two relationships, GR-QTST is therefore also strongly connected to instanton theory
and predicts an effective action similar to $\phi_\text{inst}$
at all values of $\tau$.
This is a positive result as instanton theory is rigorously derived \cite{GoldenGreens}
and gives accurate rate predictions for this type of system. \cite{AsymSysBath}
In liquids, where many diabatic crossings exist, both the steepest-descent approximation to the GR-QTST formula
and instanton theory will fail.
However GR-QTST should still be applicable to liquids and is expected to sample all the crossing seams correctly.

\subsection{One-dimensional anharmonic model}

We have given an analytical proof that GR-QTST is exact for the crossed linear system in Appendix \ref{sec:linear}\@.
To test the new method on an anharmonic system,
we study a one-dimensional model of the dissociation of a molecule via an electron transfer,
which has been employed to benchmark rate calculations in previous work \cite{nonoscillatory,Lawrence2018Wolynes}.
The potential-energy surfaces are defined as
\begin{align}
\label{AnharmonicSys}
\begin{split}
&V_0(x)=\thalf m\omega^2 x^2
\\
&V_1(x)=D \, \eu{-2\alpha(x-x_0)}-\epsilon.
\end{split}
\end{align}
The parameters are specified in reduced units such that $\hbar=1$, mass $m = 1$ and frequency $\omega = 1$.
We study the system with parameters $\epsilon=0$, $D$ = 2, $\alpha$ = 0.2 and $x_0$ = 5.
This is a non-trivial test case because
$V_-(x)$ has a nonlinear dependence on $x$
such that $s(x_+)$ appearing in \eqn{Ev} is not always equal to the crossing point, $x^\ddag$.

For comparison with GR-QTST, we also present results from exact, classical, semiclassical and Wolynes theories.
The exact rate is found using a wave-function expansion of \eqn{exactE} using the known eigenstates of the harmonic and Morse oscillators.\cite{Bunkin1973Morse,Muendel1984morse}
Semiclassical instanton rates were computed using the formula presented in \Ref{GoldenRPI}.
The classical rate for this anharmonic system is given by Landau-Zener formula \cite{Landau1932LZ,Zener1932LZ}
in the golden-rule limit \cite{Nitzan,nonoscillatory},
\begin{align}
\label{ClLZ}
k_{\textrm{LZ}}(T)=\frac{\Delta^2}{\hbar} \frac{\sqrt{2 \pi \beta m\omega^2} }{|\del V_0(x^{\ddag})-\del V_1(x^{\ddag})|} \, \eu{-\beta V_0(x^{\ddag})} \,,
\end{align}
where $x^{\ddag}$ denotes the crossing point.
This formula can be obtained from the classical limit of the instanton rate \cite{GoldenGreens}
and takes into account that 
reactive trajectories might have positive or negative values of momenta and will contribute equally.
\cite{Nitzan}

\begin{figure}
	\includegraphics{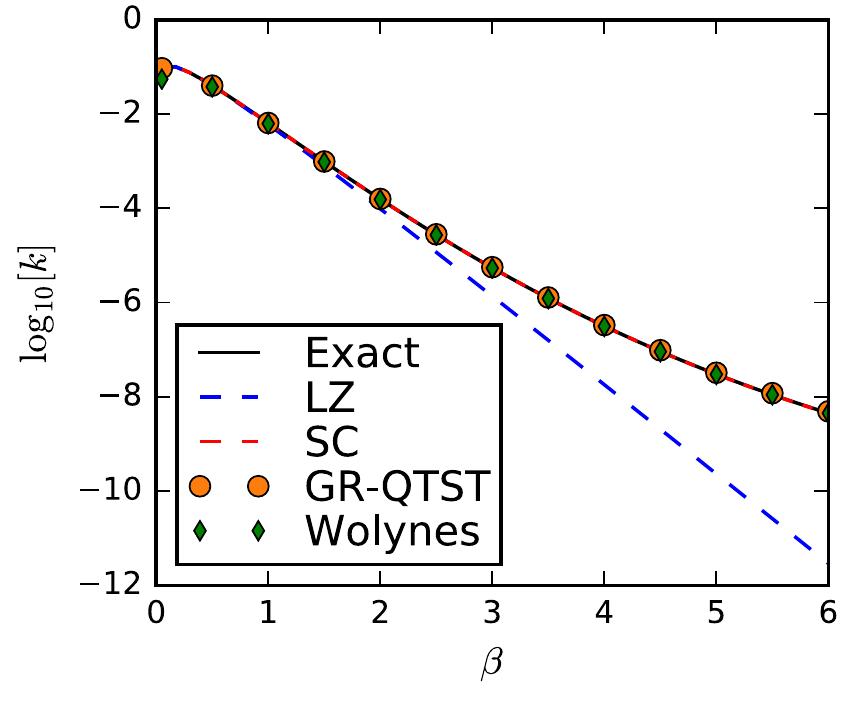}
	\caption{Rate constants calculated for the one-dimensional anharmonic system defined in \eqs{AnharmonicSys}. 
		GR-QTST rates obtained using \eqn{grqtst} are compared against exact, \eqn{exactE},
		semiclassical instanton (SC), \eqn{instphi}, %
		classical Landau-Zener (LZ), \eqn{ClLZ}, and Wolynes, \eqs{Wolynes}, theories. 
		$N=100$ was used for all path-integral methods.
		}
	\label{fig:anharmonic}
\end{figure}

As shown in Figure \ref{fig:anharmonic}, the GR-QTST method is accurate at a wide range of temperatures.
In the high temperature limit, it tends rigorously to the exact result,
as do classical Landau-Zener and the semiclassical instanton theories.
For example, at $\beta=0.005$, the GR-QTST method demonstrates very small error of about 0.2\% in the rate constant.
At lower temperatures, such as at $\beta=6$, it still gives good predictions but shows slightly higher error of 8\%.
Wolynes theory is actually more accurate than GR-QTST for this system at lower temperatures.
However, at higher temperatures, Wolynes theory fails to reproduce the classical limit,
and demonstrates an error of over 41\% at $\beta=0.005$.
In the classical limit, all beads should collapse to a point at the crossing between the two diabatic surfaces. \cite{nonoscillatory}
However, because the integrand of Wolynes theory does not contain a constraint in the path integral,
it samples many other configurations as well which contaminate the result \cite{GoldenRPI} (see \fig{FreeEnergy}). 
Although this may be seen as a rather minor problem here, its unphysical behaviour in the classical limit could have much more drastic consequences in complex systems.

\section{Multidimensional generalization}
\label{sec:multidim}

In this section we generalize the theory presented above 
to treat a system with $f$ nuclear degrees of freedom, $\mathsf{x} = (x_1,\dots,x_f)$.
Each electronic state is thus described by the Hamiltonian
$\op{H}_n = \sum_{j=1}^f \op{p}_j^2/2m + V_n(\op{\mathsf{x}})$.
We have mass-scaled each degree of freedom such that they have the same effective mass, $m$.

\subsection{Definition of the multidimensional theory}

As before, we use generalized coordinates, $\mathsf{q}=\{q_1,\dots,q_f\}$.
The first one is defined similarly to the one-dimensional case as %
\begin{align}
	q_1(\mathsf{x}) = V_-(\mathsf{x}_+) + \del V_-(\mathsf{x}_+) \cdot (\mathsf{x} - \mathsf{x}_+),
\end{align}
where $\mathsf{x}_+=\thalf(\mathsf{x}'+\mathsf{x}'')$ and $\mathsf{x}'=\mathsf{x}(0)$ and $\mathsf{x}''=\mathsf{x}(\tau)$.
The other generalized coordinates are chosen to be orthogonal to $q_1(\mathsf{x})$ and defined as %
\begin{align}
	q_k(\mathsf{x}) = \mathsf{v}_k \cdot (\mathsf{x} - \mathsf{x_+}), \qquad \text{for } k = {2,\dots,f},
\end{align}
where $\mathsf{v}_k$ form a set of vectors orthogonal to each other as well as to $\del V_-(\mathsf{x}_+)$.

Using the Einstein summation rule, the Lagrangian for each electronic state can be written
$L_n = \half m\dot{x}_j\dot{x}_j + V_n(\mathsf{x})$,
and its derivatives are
\begin{align}
	\pder{L_n}{\dot{q}_k}
	&= m \pder{x_j}{q_{k}} \pder{x_j}{q_{k'}} \dot{q}_{k'} 
	\\
	\pder{L_n}{q_k}
		&= m \dot{x}_j \der{}{t} \left(\pder{x_j}{q_k}\right)
	+ \pder{V_n}{x_j}\pder{x_j}{q_k}.
\end{align}
The Jacobian matrix for the transformation, $\mathsf{J}$, has elements $[\mathsf{J}]_{kj} = \pder{q_k}{x_j}$ 
and the inverse transformation is given by $\pder{x_j}{q_k} = [\mathsf{J}^{-1}]_{kj}$.

The thermodynamic energy functional for a path, $\mathsf{x}(t)$, is given by
\begin{align}
	\bar{E}_n^\text{th}[\mathsf{x}(t)] 
	&= -\frac{1}{2\tau_n}\left. \pder{L_n}{\dot{q}_k}q_k \right|_0^{\tau_n} 
	\nonumber\\&\quad+ \frac{1}{\tau_n} \int_0^{\tau_n} \left[ \half \der{}{t}\left(\pder{L_n}{\dot{q}_k}\right) q_k + V_n(\mathsf{x}) \right] \rmd t 
\end{align} 
and by following the same steps as in \secref{theory}, the virial energy functional is found to be
\begin{align} 
	\bar{E}_n^\text{v}[\mathsf{x}(t)]  
	&= \frac{1}{\tau_n} \int_0^{\tau_n} \left[ \half \pder{L_n}{q_k} q_k + V_n(\mathsf{x}) \right] \rmd t,
\end{align}  
where $\pder{L_n}{q_k} q_k=\pder{V_n}{x_j} \big( x_j - s_j(\mathsf{x}_+) \big)$ and
$\mathsf{s}(\mathsf{x}_+) = \mathsf{x}_+ - \pder{\mathsf{x}}{q_1} V_-(\mathsf{x}_+)$.
This was derived using the Euler-Lagrange equation, \eqn{EL}, for each degree of freedom.
As before, the boundary term can be shown to vanish for the instanton trajectory
because the following relations are obeyed at the end points:
${\qinst'}_1={\qinst''}_{1}=0$ and ${{}\dotqinst'}_k={{}\dotqinst''}_k=0$ for $k=2,\dots,f$.
This follows from the proof in \Ref{GoldenGreens} that the instanton changes state on the crossing seam
and at this point has velocity perpendicular to the seam, which is in the $q_1$ direction only.

All other formulae given in \secref{ringpolymer} and \secref{PIMCimplement}
are easily generalized by replacing the scalar $x$ with the vector $\mathsf{x}$ %
and $\Lambda$ by $\Lambda^f$.
If we also allow the coupling to become dependent on position, %
the prefactor $\Delta^2$ should be replaced by $\Delta(\mathsf{x}')\Delta(\mathsf{x}'')$, and moved inside the integral of \eqn{phi} \cite{nonoscillatory}.

The GR-QTST rate tends correctly to the classical limit also for multidimensional systems.
This is because %
for short paths, which occur in the classical limit, $q_1$ is simply $V_-(\mathsf{x}_+)$.
This gives $\sigma_{\tau}(\mathsf{x})=\beta V_-(\mathsf{x})$,
which is the required constraint in this limit, \cite{nonoscillatory}
similarly to the one-dimensional case outlined in Appendix~\ref{sec:classical}.

\subsection{Application to system-bath model}
\label{sec:sysbath}

The new GR-QTST method is applied to a multidimensional system-bath model, and the rates obtained are compared against existing methods. 
The potential-energy surfaces of the system-bath model are defined as follows,
\begin{align}
V_n(\mathsf{x})=V_n(x_1)+V_\text{b}(x_1,\dots,x_f),
\end{align}
where
\begin{align}
V_0(x_1)&=\thalf m\Omega^2(x_1+\xi)^2
\\
V_1(x_1)&=\thalf m\Omega^2(x_1-\xi)^2-\epsilon,
\end{align}
and $V_\text{b}$ describes the bath which couples $x_1$ to the other degrees of freedom:
\begin{align}
V_\text{b}(x_1,\dots,x_f) = \sum_{j=2}^f \thalf m \omega_j^2 \left(x_j-\frac{c_j}{m \omega_j^2}x_1 \right)^2.
\end{align}

The bath is defined by the Ohmic spectral density, $J(\omega)=m \gamma \omega \, \eu{-\omega/\omega_\text{c}}$,
where $m\gamma$ is the friction coefficient and $\omega_\text{c}$ is the cut-off frequency.
The bath is discretized into $f-1$ modes as \cite{Weiss}
\begin{align}
J(\omega)=\frac{\pi}{2}\sum_{j=2}^f\frac{c_j^2}{m \omega_j}\delta(\omega-\omega_j),
\end{align}
where \cite{RPMDrate}
\begin{subequations}
\begin{align}
\omega_j&=-\omega_\text{c} \, \ln\left(\frac{j-3/2}{f-1}\right)
\\
c_j&=m \omega_j \sqrt{\frac{2 \gamma \omega_\text{c}}{\pi (f-1)}}.
\end{align}
\end{subequations}
The one-dimensional system ($f=1$) has no bath and is thus frictionless,
whereas all other systems are dissipative with the same damping, $\gamma$.
Note that the Hamiltonian of this system-bath model is equivalent by unitary transformation to that of a spin-boson model with Debye spectral density \cite{Thoss2001hybrid}.

The system considered in our numerical tests
has reorganization energy $\lambda\equiv2 m \xi^2 \Omega^2$ = $\SI{80}{\kilo\cal\per\mol}$
and the bias, $\epsilon$, is varied from $0$ to $2\lambda$.
The system frequency is $\Omega/2\pi c = \SI{500}{\per\cm}$ and the bath cut-off frequency $\omega_\text{c}/2\pi c=\SI{500}{\per\cm}$,
where here $c$ is the speed of light in vacuum.
The friction coefficient is defined by $\gamma \hbar=0.001\,\hartree$, where $\hartree$ is the Hartree energy.
This is a medium strength friction, strong enough to ensure that the reaction dynamics is incoherent
but not too large as to significantly reduce the rate. \cite{Cline1987nonadiabatic,*Onuchic1988rate,Topaler1996nonadiabatic,Rips1995ET}
Rates are computed at a temperature of $T=\SI{300}{\kelvin}$.
Note that the rate for this system is independent of mass, $m$.

The classical rate constant, \eqn{cl}, of the system-bath model in the golden-rule limit is given by Marcus theory as \cite{Marcus1985review}
\begin{align}
	\label{ClMarcus}
	k_{\textrm{MT}} = \frac{\Delta^2}{\hbar}\sqrt{\frac{\pi \beta}{\lambda}} \, \eu{-\beta(\lambda-\epsilon)^2/4 \lambda}.
\end{align}
The exact Fermi's golden-rule rate was
calculated using the analytical path-integral expression \cite{Weiss} for the flux-flux correlation function \cite{Miller1983rate}
and numerically integrated to the first plateau.
\cite{Bader1990golden}

In previous work, we have shown that instanton theory gives excellent predictions for the rate of the reaction in this model \cite{AsymSysBath}.
However, the ring-polymer instanton method is not numerically applicable in the inverted regime
because $\tauinst$ would fall outside of the range [0, $\beta \hbar$].
Nonetheless, one can use analytically obtained instantons to predict semiclassical rates in the inverted regime for a finite number of beads.
To obtain an analytical expression for the instanton rates,
we consider a coordinate transform of the multidimensional system-bath Hamiltonian to a spin-boson Hamiltonian. \cite{Garg1985spinboson,Thoss2001hybrid}
$\Omega_j$ are the normal-mode frequencies and $\xi_j$ are the displacements of the harmonic oscillators that 
result from this coordinate transformation. 
For $N$ beads, the semiclassical instanton rate constant is given by the following formula \cite{GoldenGreens}:
\begin{align}
\label{SCNrate}
	k_\text{SC}^{(N)}=\sqrt{2 \pi \hbar} \, \frac{\Delta^2}{\hbar^2}\left[-\frac{\rmd ^2\Sinst^{(N)}}{\rmd \tau^2}\right]_{\tau=\tauinst}^{-\frac{1}{2}} \eu{-\Sinst^{(N)}/\hbar},
\end{align}
where $\Sinst^{(N)}$ is the action corresponding to an instanton trajectory discretized with $N$ ring-polymer beads for $\tau=\tauinst$, and is defined as
\begin{align}
	\Sinst^{(N)} = -\epsilon \tau_1 - \sum_{j=1}^{f} 8m\xi_j^2\frac{\Upsilon_{j}^{(0)} \Upsilon_{j}^{(1)}}{\Upsilon_{j}^{(0)} -\Upsilon_{j}^{(1)}},
\end{align}
where $\Upsilon_{j}^{(n)}=\frac{N_n \sinh{(\zeta_{n,j}/N_n)}}{2 \tau_n \sinh \zeta_{n,j}}\left[\cosh\zeta_{n,j}-1\right]$
and $\zeta_{n,j}=2 N_n \arsinh{(\Omega_j \tau_n/2 N_n)}$.
\cite{Kleinert}
In our calculations, we chose to distribute an equal number of beads on the reactant and product surfaces,
i.e. $N_0=N_1=N/2$.
To compute the rate in \eqn{SCNrate}, one chooses $\tauinst$ numerically as the value of $\tau$ which solves
$\der{}{\tau} \Sinst^{(N)}=0$.

To calculate the GR-QTST rates, the Monte Carlo scheme we proposed in \secref{PIMCimplement} is employed.
The importance sampling scheme turns out to be extremely efficient for this problem
and we are able to get an extremely small statistical error of less than 1\% with about $10^4$ random samples for this system-bath model with 8 degrees of freedom.

\begin{figure}
\includegraphics{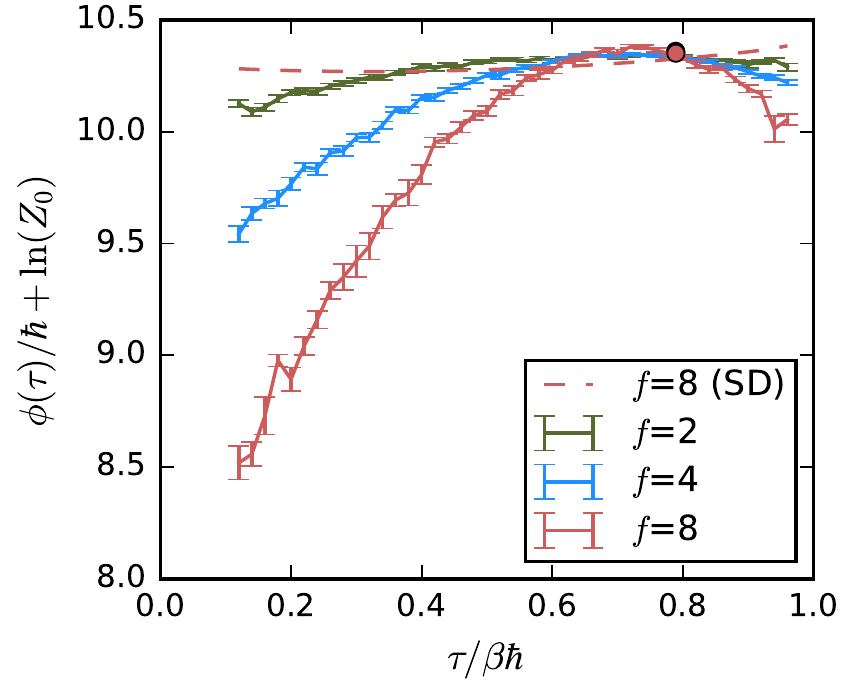}
	\caption{
		The dependence of the effective action on $\tau$
		for the system-bath model
		with $\epsilon/ \lambda = 0.5$ and in each case, a different number of dimensions, $f$.
		All calculations were performed using $N=50$.
		GR-QTST results, computed using \eqn{grqtst}, are shown with statistical error bars.
		The result of the steepest-descent (SD) approximation, \eqn{sdapprox}, is included only for $f=8$
		as for all multidimensional ($f \geq 2$) cases, the SD approximations to $\phi(\tau)$ are approximately the same.
		The effective actions of the instanton method, $\phi_{\textnormal{inst}}$, are marked with dots at $\tau=\tauinst$ for each case,
		although these lie almost exactly on top of each other.
		All results are normalized by the partition function $Z_0$ to ease comparison with the frictionless results shown in \fig{phi1d}.
	}
	\label{fig:phi}
\end{figure}

In \fig{phi}, we show the form of the effective action, $\phi(\tau)$,
to investigate its behaviour on adding extra degrees of freedom.
In Appendix \ref{sec:linearharmonic}
we examined the effect of adding uncoupled bath modes to the linear crossing model.
In this case where the bath is coupled to the system, the behaviour is similar in many ways.
The effect on $\phi_\text{SD}(\tau)$ is to simply shift it to slightly higher values but it remains approximately flat.
However, adding extra degrees of freedom to the model has a larger effect on $\phi(\tau)$,
which gets increasingly curved around its maximum with each additional degree of freedom.
The value of $\tau^*$ chosen to compute the GR-QTST rate is the one which maximizes these curves,
and this value is found at approximately the same value as for the instanton, $\tauinst$.
At this point, the value of $\phi(\tau^*)$ is very similar to that of the instanton, $\phi_\text{inst}$,
and thus GR-QTST will predict a rate similar to that of instanton theory as we wanted.

\begin{figure}
	\includegraphics{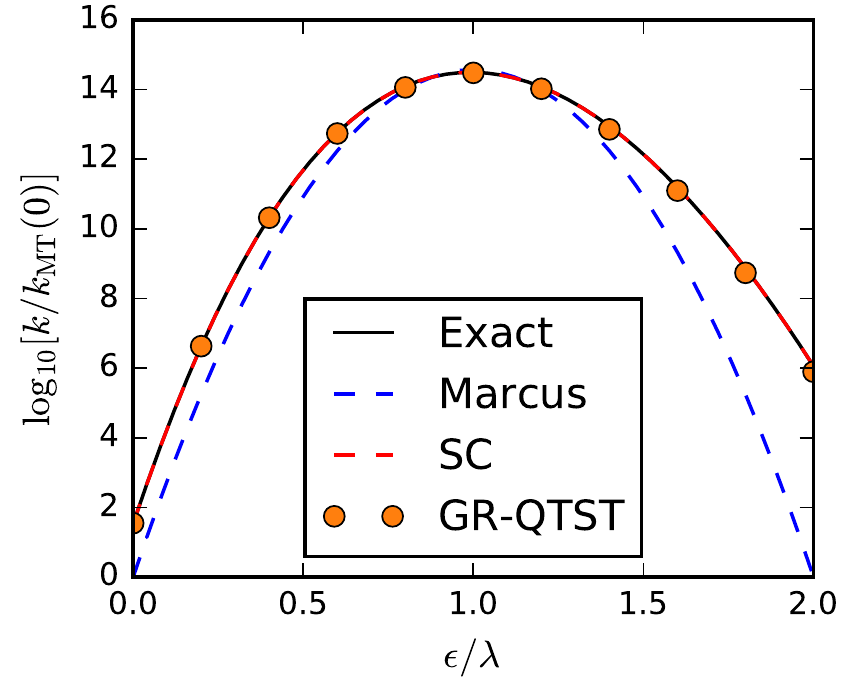}
	\caption{
		The calculated rate constants are shown as a function of bias, $\epsilon$,
		for the 8-dimensional system-bath model.
		The various approaches used are
		exact Fermi's golden rule \cite{fermiGR},
		classical Marcus theory, \eqn{ClMarcus},
		semiclassical instanton (SC), \eqn{SCNrate}, and GR-QTST\@.
		Wolynes theory is equal to SC for this case but is only numerically applicable in the normal regime, $\epsilon<\lambda$.
		The reference, $k_\text{MT}(0)$, is the classical Marcus theory rate constant for $\epsilon=0$.
		SC rates were computed analytically using $N \to \infty$,
		whereas the GR-QTST rates were obtained with $N=100$ ring-polymer beads.
	}
	\label{fig:spinboson}
\end{figure}

Figure \ref{fig:spinboson} shows the results for this multidimensional system-bath model 
for varying degrees of bias, $\epsilon$. 
It is seen that the semiclassical and GR-QTST methods both give excellent predictions for the rate both in the normal and inverted regimes, where $\epsilon>\lambda$.
Because in this case the reactant and product surfaces are harmonic,
Wolynes theory gives the same result as semiclassical instanton theory.
However, neither Wolynes theory nor ring-polymer instanton theory is numerically applicable to study nonadiabatic reactions in the inverted regime. %
It is therefore particularly interesting that
GR-QTST gives accurate predictions even deep into this regime. %

In the inverted regime,
the optimal value of $\tauinst$ which makes the action stationary is greater than $\beta\hbar$.
According to the ansatz of GR-QTST the rate should be evaluated at $\tau^*=\beta \hbar$, which gives the maximum $\phi(\tau^*)$ which can be obtained.
As shown in Figure \ref{fig:minimalS}, because $S_\tau(\cmin{\mathbf{x}})$ is quite flat with respect to $\tau$ and in fact exactly so in the linear case,
the effective actions sampled will be similar to the action of the analytic instanton solution
and thus the rate prediction, like that of instanton theory, will be a good approximation to the exact result.

We did not have to rely on analytic continuation of the function $\phi(\tau)$ outside the range $[0,\beta\hbar]$
in order to obtain this result
because the GR-QTST rate is approximately independent of the value of $\tau$.
However, because it can be curved, especially with many degrees of freedom,
it may be possible to obtain a better prediction using extrapolation
to locate the optimal splitting $\tau^{*}$,
as has been done previously from Wolynes theory in \Ref{Lawrence2018Wolynes}.
As our function is flatter than the unconstrained form, the extrapolation will be more robust and less likely to give an unphysical prediction.

\begin{table*}
\begin{ruledtabular}\begin{tabular}{ llllll }
\multirow{5}{*}{ }$\epsilon/\lambda$ & method & $f=1$ & $f=2$ & $f=4$ & $f=8$ \\
\hline
\multirow{4}{*}{0.0} & exact & 37.81 & 33.65 & 33.92 & 33.97\\
&SC ($N\to\infty$) & 37.68 & 33.55 & 33.81 & 33.86\\
&SC ($N=100$) &37.62 & 33.50 & 33.76 & 33.83\\
&GR-QTST ($N=100$) & 39.11(3) & 34.93(5) &34.90(8) & 35.27(9)\\ \hline
\multirow{4}{*}{0.5} & exact &5.987 & 5.738 & 5.753 & 5.757\\
&SC ($N\to\infty$) &5.980 & 5.729 & 5.747 & 5.750\\
&SC ($N=100$) &5.961 & 5.713 & 5.737 & 5.764\\ 
&GR-QTST ($N=100$) &6.12(2) & 5.81(3) & 5.83(3) & 5.73(3)\\ \hline
\multirow{4}{*}{1.0} & exact  &0.8334 & 0.8362  & 0.8359 & 0.8359 \\
&SC ($N\to\infty$) &0.8337 & 0.8363 & 0.8363 & 0.8361\\
&SC ($N=100$) &0.8337 & 0.8364 & 0.8364 & 0.8362\\
&GR-QTST ($N=100$) &0.826(3) & 0.838(3) & 0.829(3) & 0.807(4)\\ \hline
\multirow{4}{*}{1.5} & exact  & 16.63 & 16.25  & 16.27 & 16.28 \\
&SC ($N\to\infty$) &16.64& 16.25 & 16.28 & 16.28\\
&SC ($N=200$) &  16.63 & 16.25 & 16.27 & 16.28\\
&GR-QTST ($N=200$) &  15.73(4) & 15.03(8) & 14.57(9) & 13.51(9)\\ \hline
\multirow{4}{*}{2.0} & exact  & $1.176 \times 10^{6}$ & $1.129 \times 10^{6}$  & $1.134 \times 10^{6}$ & $1.134 \times 10^{6}$ \\
&SC ($N\to\infty$) &$1.178\times 10^{6}$& $1.127 \times 10^{6}$ & $1.134 \times 10^{6}$ & $1.134 \times 10^{6}$\\
&SC ($N=200$) &  $1.174 \times 10^{6}$ & $1.126 \times 10^{6}$& $1.131 \times 10^{6}$ & $1.132 \times 10^{6}$\\
&GR-QTST ($N=200$) &  $1.001(2) \times 10^{6}$ & $0.925(3) \times 10^{6}$& $0.883(8) \times 10^{6}$ & $0.790(9) \times 10^{6}$\\ %
\end{tabular}\end{ruledtabular}
\caption{Results of exact, semiclassical instanton (SC), \eqn{SCNrate}, and GR-QTST calculations
on the system-bath model defined in the text for
different values of bias, $\epsilon$, and number of degrees of freedom, $f$.
Exact rate constants were computed by numerically integrating over the analytic expression for the flux-flux correlation function
as far as the first plateau.
In the table, we present the quantum factor $k/k_\text{MT}$, where the Marcus theory rate constant was calculated using \eqn{ClMarcus}.
The GR-QTST results are reported along with the Monte Carlo sampling error, which is given for the last digit in parentheses.}
\label{tab:rates}
\end{table*}

In order to analyse the accuracy of the various methods in more detail,
the calculated rate constants are given in \tref{rates}.
All results are reported relative to the classical Marcus theory rate, which is independent of $f$.

For the frictionless one-dimensional system,
the exact rate is technically not defined as the quantum dynamics is coherent.
Instanton theory and GR-QTST, like other transition-state theories, cannot describe this behaviour
and thus for comparison, we give the value obtained by integrating the flux-flux correlation function up to the first plateau.
We have checked that the friction of our system is strong enough to ensure that the revivals are destroyed and the exact rate is well defined for the multidimensional systems.
About 8 degrees of freedom are needed to converge these rates to three significant figures.
The exact calculations for this system indicate that the Fermi's golden-rule rates
decrease slightly because of the coupling to the bath modes. 
This is a well-known effect in dissipative systems, in which friction hinders tunnelling. \cite{Caldeira1983dissipation,Weiss,Mak1991spinboson,Topaler1996nonadiabatic}

As has been observed in previous work, \cite{AsymSysBath}
the semiclassical instanton approximation has excellent accuracy in predicting the rates of the spin-boson model.
This is
due to the fact that as the potentials are harmonic, the steepest-descent approximation is exact for all nuclear degrees of freedom.
By comparison with semiclassical calculations with a finite number of beads,
we also show that the discretization error is very small.
In the normal regime, it is seen that $N=100$ is enough to converge the results to three significant figures,
whereas in the inverted regime, because of particularly large tunnelling effects, one needs $N=200$.

Rates obtained by GR-QTST are then compared with the instanton method for the same number of ring-polymer beads, $N$.
It is seen that GR-QTST makes an error of less than 4\% in each case in the normal regime, where $\epsilon<\lambda$.
However, deep in the inverted regime, where $\epsilon/\lambda=2$, the new method demonstrates slightly higher error of about 30\%,
but nonetheless continues to give a correct order-of-magnitude estimate of the rate
even though the quantum effects are greater than a million.

\section{Discussion}
\label{sec:discussion}

In this paper, we have described a new quantum transition-state theory
for the calculation of nonadiabatic rates in the golden-rule limit.
The current work only presents applications to low-dimensional model systems
for which Monte Carlo importance sampling is simple and accurate to apply.
However, in order to employ the new approach with atomistic simulations of electron-transfer reactions in solution,
different sampling schemes will be required.
Due to the quantum-classical correspondence of the path-integral formalism, 
many of the molecular dynamics sampling techniques developed for classical statistical mechanics
can be used in an extended ring-polymer space to evaluate 
the imaginary-time path integration in a numerically efficient manner \cite{Chandler+Wolynes1981,Parrinello1984Fcenter,Marx1996PIMD}.
For instance, the GR-QTST rate could be computed with 
biased path-integral molecular dynamics
using a thermodynamic integration along the reaction coordinate $\sigma_\tau(\mathbf{x})$. 
We shall explore these possibilities in future work.

In contrast to Wolynes theory, \cite{Wolynes1987nonadiabatic} the new method rigorously tends to the correct classical limit.
For a system of crossed linear potential-energy surfaces uncoupled to a harmonic oscillator bath,
it gives the exact Fermi's golden-rule rate.
Furthermore, due to its connection to instanton theory, 
it also gives a good order-of-magnitude prediction for more general systems for a wide range of temperatures. 
It is particularly interesting that it is able to go beyond ring-polymer instanton theory
and make accurate predictions for rates deep in the inverted regime.

Note that other forms of the ansatz, \eqs{approx}, might also exist.
We originally considered $Z^\ddag\approx\int_0^{\beta \hbar}{\eu{-\phi(\tau)/\hbar} \, \rmd\tau}$,
which is similar to a rate formulation in terms of the Kubo transform \cite{Kubo,Yamamoto1960rate}.
This was found to work well for one-dimensional systems, where $\phi(\tau)$ is almost flat,
but was not as accurate for the case of multidimensional systems. %
In future work, we shall investigate variations of the constraint functional which could be used with an alternative ansatz.

Previous QTST approaches designed to give the adiabatic rate on a single Born-Oppenheimer surface
have used a constraint based on dividing surfaces in ring-polymer configuration space
\cite{Gillan,Voth+Chandler+Miller,Mills1997QTST,RPInst,Hele2013QTST}.
This introduces the numerically challenging problem of locating the optimal dividing surface in each case
which makes them difficult to apply in general to complex systems.
One thus uses RPMD which is independent of this choice. \cite{RPMDrate,RPMDrefinedRate,RPMDprotonTransfer}
In contrast to these approaches,
the GR-QTST formulation
employs a completely different constraint functional based on energy conservation
which does not need to be variationally optimized.
Because of this, we do not require dynamical methods like RPMD in this case.

This work also points towards
the development of a general nonadiabatic quantum transition-state theory
which would be valid beyond the golden-rule approximation.
In this more general regime, the exact rate constant can still be written
as two Boltzmann factors separated by flux operators
with an energy conservation requirement \cite{nonoscillatory}.
The only difference is that hops between surfaces are also allowed within the Boltzmann operators.
The imaginary-time path integrals for such problems are easily evaluated using matrix algebra \cite{Alexander2001diabatic}.
If an appropriate energy constraint functional can be devised,
such a method would be able to describe a wide range of chemical reactions 
from those occurring in the golden-rule limit, where $\Delta\rightarrow0$,
to those in the Born-Oppenheimer regime where $\Delta$ is large.

Some progress in this direction has been achieved already,
\cite{Cao1995nonadiabatic,Cao1997nonadiabatic,*Cao1998erratum,Schwieters1998diabatic,*Schwieters1999diabatic,Zhu1994ZN,*Zhu1995ZN}
but these methods
do not include energy constraints and are thus not directly comparable to our GR-QTST approach.
The kinetically-constrained RPMD approach (KC-RPMD) \cite{Menzeleev2014kinetic,Kretchmer2016KCRPMD,kretchmer2018fluctuating},
does employ an energy constraint to perform the path-integral sampling.
The constraint is however applied only on the centroid which 
means that such a constraint surface will not pass through the instanton for asymmetric barriers
and the instanton configuration will not be included in the path-integral sampling.
KC-RPMD is a dynamical method, which makes use of real-time trajectories
to correct for recrossing effects,
and is designed to predict rates for all coupling strengths, $\Delta$. %
Dynamics are performed in an extended space including
an auxiliary variable which reports on the formation of hops in the ring polymer.
Due to all of these differences, it is hard to see exactly how KC-RPMD and GR-QTST are related.
However, we can at least make some numerical comparisons.
In \Ref{Kretchmer2016KCRPMD}, the authors report results for a symmetric system-bath model (called System A3).
\footnote{%
We have checked that these systems are in the golden-rule limit by comparing our Fermi's golden-rule calculations with
the exact results of \Ref{Topaler1996nonadiabatic}, which are essentially identical.
}
For this system with intermediate friction, $\gamma/\Omega=1$,
GR-QTST gives $k/k_\text{cl}=0.048(4)$,
where $k_\text{cl}$ is the classical adiabatic TST rate \cite{Topaler1996nonadiabatic}.
This is in good agreement with the exact result, $0.053$,
whereas the KC-RPMD result reported in that paper is about 5 times too large.
For the same symmetric system, in the weaker dissipative regime, $\gamma/\Omega\le 0.1$,
where nuclear coherence effects become significant, \cite{Cline1987nonadiabatic,*Onuchic1988rate,Topaler1996nonadiabatic}
neither KC-RPMD nor GR-QTST is able to describe the 
enhancement of the exact rate. %
Instead, the GR-QTST 
rate becomes independent of friction,
which is in agreement with the result obtained by integrating the analytic flux-flux correlation function up to the first plateau.
The KC-RPMD rate deep in the inverted regime ($\epsilon=0.236\,\hartree$) for System B of \Ref{Menzeleev2014kinetic}
is also reported to be too large compared with the exact result by almost an order of magnitude.
For the same system, the GR-QTST method gives $k/k_\text{MT}=8.0(2) \times 10^{6}$,
demonstrating a smaller error compared to the exact result, $1.4 \times 10^{7}$.

Other approaches extract the quantum rate constant from a real-time dynamical simulation of the nonadiabatic reaction.
Examples include those based on extensions of RPMD 
to employ the mapping formalism \cite{Meyer1979nonadiabatic,Stock1997mapping,mapping,Ananth2013MVRPMD,Duke2016Faraday,Pierre2017MVRPMD,Chowdhury2017CSRPMD},
surface hopping  \cite{Shushkov2012RPSH,Lu2017RPSH,Shakib2017RPSH}
or explicit electron dynamics in the position representation \cite{Menzeleev2011ET,Kretchmer2013ET} as well a related instanton theory \cite{Shushkov2013instanton}.
There are also mixed quantum-classical dynamical methods \cite{Stock2005nonadiabatic,Kapral2015QCL} based on
linearized and partially linearized path-integral simulations \cite{Wang1999mapping,Kelly2012mapping,Huo2013PLDM},
Bohmian dynamics \cite{Curchod2013Bohmian}
or the exact factorization of the time-dependent electron-nuclear wave function \cite{Agostini2013MQC}.
An isomorphic Hamiltonian 
can be used to include quantum nuclear effects in nonadiabatic trajectory simulations \cite{Tao2018isomorphic}.
It is expected that in an accurate nonadiabatic RPMD simulation, the ring polymer will pass through transition-state configurations
similar to those sampled in the GR-QTST ensemble.
We therefore expect that our study of GR-QTST will be of use in understanding and improving the dynamical methods discussed above.

\section{Acknowledgements}
We are grateful for the ETH Zurich Research Grant that supports M.J.T. in his doctoral studies. %
The authors also acknowledge support from the Swiss National Science Foundation through the NCCR MUST (Molecular Ultrafast Science and Technology) Network. 

\appendix
\section{Connection to the classical limit}
\label{sec:classical}

As shown by Feynman, \cite{Feynman}
the classical, high-temperature, limit of a path integral can be obtained by noting that paths will collapse to a single point,
$x(t)\rightarrow x$ and $\dot{x}\rightarrow0$,
such that functionals become functions of a position and path integrals become configuration integrals.
The classical limit of the constrained effective action is thus given by
\begin{align}
	\label{phicl}
	\eu{-\phi(\tau)/\hbar}
	&= \sqrt\frac{m}{2\pi\beta\hbar^2} \int \eu{-\tau_0V_0(x)/\hbar-\tau_1V_1(x)/\hbar} \, \delta(\sigma_\tau(x)) \, \rmd x.
\end{align}

The constraint functional also simplifies due to the collapsed path.
Using $x_+ = x$, \eqn{q} simplifies to give $q(x) = V_-(x)$ and hence
\begin{align}
	\sigma_\tau(x) &= \frac{2}{3}\beta\left( \half\left(\pder{V_0}{q}-\pder{V_1}{q}\right) q(x) + V_0(x) - V_1(x) \right)
	\nonumber\\
	&= \beta V_-(x) \, .
	\label{sigmacl}
\end{align}
Therefore the integral in \eqn{phicl} is independent of $\tau$.
In this way, one recovers the correct classical rate expression \cite{nonoscillatory}
\begin{align}
	\label{cl}
	k_\text{cl} Z_0 = \frac{2\pi\Delta^2}{\hbar} \sqrt\frac{m}{2\pi\beta\hbar^2} \int \eu{-\beta V_0(x)} \, \delta[V_0(x) - V_1(x)] \, \rmd x \, .
\end{align}

The necessity of the factor of $\frac{2}{3}$ is now seen clearly.
As is known from the virial theorem,
in the classical limit
the contribution to the total energy from the kinetic energy is half that of the potential energy
and this factor was necessary to normalize the term.
This general classical limit is also valid in the Marcus inverted regime.

\section{Analytical solution for linear system}
\label{sec:linear}

In this appendix, we consider
the linear system defined by
\begin{align}
	\label{Vlinear}
	V_n(x) &= \kappa_n x
\end{align}
with $\kappa_0 \neq \kappa_1$.
If the signs of $\kappa_0$ and $\kappa_1$ are different,
this system describes electron transfer in the normal regime.
However, if the signs are the same, the reaction is in the inverted regime.
The following proof is valid for both cases.

The form of \eqn{Vlinear} could be trivially generalized to $V_n(x)=V^\ddag+\kappa_n(x-x^\ddag)$.
This would make only minor differences to the following equations,
e.g.\ by including an extra term $\beta\hbar V^\ddag$ in the action.
Using the definition of the $q$ coordinate, it is clear that the rate expression is independent of $x^\ddag$.
We therefore use the simpler expression \eqn{Vlinear} to keep the following equations as compact as possible.

The derivation can be carried out analytically because
The instanton and functionals of its trajectory are known for this system. \cite{GoldenGreens}
The total action of a ring formed of two classical trajectories is %
$\tilde{S}(x_+,x_-,\tau) = \tilde{S}_0(x_+,x_-,\tau_0) + \tilde{S}_1(x_+,x_-,\tau_1)$,
where we employ the 
following change of coordinates for the end points: 
\begin{align}
	x_+ &= \thalf (x'+x'')
	&
	x_- &= x'-x''.
\end{align}
The action along a classical trajectory is known analytically: \cite{Brown1994linear}
\begin{align}
	\tilde{S}_n(x_+,x_-,\tau_n) &= \frac{m x_-^2}{2\tau_n} + \kappa_n \tau_n x_+ - \frac{\kappa_n^2\tau_n^3}{24m}. %
\end{align}
The instanton is defined at the stationary point of $\tilde{S}(x_+,x_-,\tau)$,
which in this case is given by
\cite{GoldenGreens}
$x_-=0$, $x_+=0$ and $\tauinst=-\kappa_1/(\kappa_0-\kappa_1)$.
The total instanton action is
therefore
\begin{align}
	\label{Sinst}
	\Sinst = \tilde{S}(0,0,\tauinst)
	&= -\frac{\beta^3\hbar^3\kappa_0^2\kappa_1^2}{24m(\kappa_0-\kappa_1)^2}.
\end{align}
It will also be useful to have the formula for the average potential energy, 
defined as
$\bar{V}_n[x(t)] = \frac{1}{\tau_n} \int_0^{\tau_n} V_n(x) \, \rmd t$,
along a trajectory,
\begin{align}
	\bar{V}_n(x_+,x_-,\tau_n) &= \kappa_n x_+ - \frac{\kappa_n^2\tau_n^2}{12m}. %
\end{align}

In order to perform the integral over the non-classical paths,
we follow the procedure developed by Feynman \cite{Feynman}
and write
\begin{align}
	x(t) = \tilde{x}(t) + \delta x(t),
\end{align}
where $\tilde{x}(t)$ is the classical trajectory which obeys 
the classical equations of motion %
as well as the boundary conditions, $x(0)=x(\beta\hbar)=x'$ and $x(\tau)=x''$.
The action is stationary about this trajectory.
The fluctuations can be expanded as a Fourier series
\begin{subequations}
\begin{align}
	\delta x(t) &= \sum_{k=1}^{N_0-1} \gamma_k^{(0)} \sin(k \pi t / \tau_0) && \text{for } t < \tau
	\\
	\delta x(t) &= \sum_{k=1}^{N_1-1} \gamma_k^{(1)} \sin(k \pi (t-\tau) / \tau_1) && \text{for } t>\tau
\end{align}
\end{subequations}
and we shall take the $N_n\rightarrow\infty$ limit to obtain the exact result.
Note that this discretization scheme is not the same as for the ring polymer, but all discretizations are equivalent in 
$N\rightarrow\infty$ limit.

By expanding the path in this way and performing the integrals over time,
the total action can be written
\begin{align}
	S_\tau[x(t)] &= \tilde{S}(x_+,x_-,\tau)
		+ \thalf \mathbf{\gamma_0}\cdot\mathbf{A_0}\cdot\mathbf{\gamma_0} + \thalf \mathbf{\gamma_1}\cdot\mathbf{A_1}\cdot\mathbf{\gamma_1},
\label{Sfluct}
\end{align}
where the fluctuation terms are written in terms of the Fourier coefficients,
$\mathbf{\gamma_n}=(\gamma_1^{(n)},\dots,\gamma_{N_n-1}^{(n)})$, as
\begin{align}
	\thalf \mathbf{\gamma_n}\cdot\mathbf{A_n}\cdot\mathbf{\gamma_n} &= \sum_{k=1}^{N_n-1} \frac{m (\pi k \gamma_k^{(n)})^2}{4\tau_n}.
\end{align}

The generalized coordinate is
$q(x) = (\kappa_0-\kappa_1)x$,
giving
$\pder{L_n}{q} = \frac{\kappa_n}{\kappa_0-\kappa_1}$,
such that
\begin{align}
	\bar{E}_n^\text{v}[x(t)]
	&= \frac{1}{\tau_n} \int_0^{\tau_n} \left[ \thalf \kappa_n x + V_n(x) \right] \rmd t
	\\
	&= \tfrac{3}{2}\bar{V}_n[x(t)]. %
\end{align}
Accounting for the factor of $\tfrac{2}{3}$, we get
\begin{align}
	\sigma_\tau[x(t)]
	&= \beta \big( \bar{V}_0[x(t)] - \bar{V}_1[x(\tau+t)] \big).
\end{align}
As it is defined as the difference between energy functionals, 
the constraint functional, $\sigma_\tau[x(t)]$ would be independent of an energy bias, $V^\ddag$,
and of shifting the $x$ coordinates. %
This is due to having used the generalized coordinate $q=q(x)$ to allow neglecting the boundary terms.

The constraint functional can then be written
\begin{align}
	\sigma_\tau[x(t)]/\beta &= \bar{V}_0(x_+,x_-,\tau_0) - \bar{V}_1(x_+,x_-,\tau_1)
		\nonumber\\&\quad+ \mathbf{b_0} \cdot \mathbf{\gamma_0} - \mathbf{b_1} \cdot \mathbf{\gamma_1}
	\\
	&= (\kappa_0-\kappa_1)x_+ - \frac{\kappa_0^2\tau_0^2-\kappa_1^2\tau_1^2}{12m}
	\nonumber\\&\quad
\label{constraint}		+ \mathbf{b_0} \cdot \mathbf{\gamma_0} - \mathbf{b_1} \cdot \mathbf{\gamma_1},
\end{align}
where
\begin{align}
	\mathbf{b_n} \cdot \mathbf{\gamma_n} &= \sum_{k=1}^{N_n-1} \frac{\kappa_n(1-(-1)^k) \gamma_k^{(n)}}{\pi k}.
\label{fluctconstaint}
\end{align}

Noting that changing the integration variables for the path integral to the Fourier coefficients $\gamma_k^{(n)}$
introduces a Jacobian, $J_n$,
the integral can be written
\begin{align}
	\eu{-\phi(\tau)/\hbar}
		&= J_0 \Lambda_0^{-N_0} J_1 \Lambda_1^{-N_1} \iint F_\tau(\gam{0},\gam{1}) 
		\nonumber\\&\quad
		\times G_\tau(\gam{0},\gam{1}) \, \rmd \mathbf{\gamma_0} \, \rmd \mathbf{\gamma_1},
\end{align}
where $\Lambda_n=\sqrt{2\pi\tau_n\hbar/mN_n}$.
The fluctuation terms are
\begin{align}
	F_\tau(\gam{0},\gam{1}) = \exp\left[- \frac{1}{2\hbar}\mathbf{\gamma_0}\cdot\mathbf{A_0}\cdot\mathbf{\gamma_0} - \frac{1}{2\hbar}\mathbf{\gamma_1}\cdot\mathbf{A_1}\cdot\mathbf{\gamma_1} \right]
\end{align}
and
\begin{align}
	G_\tau(\gam{0},\gam{1})
	&= \iint \eu{-\tilde{S}(x_+,x_-,\tau)/\hbar} \, \delta(\sigma_\tau) \, \rmd x_- \, \rmd x_+
	\\
	&= \sqrt\frac{2\pi\hbar}{m/\tau_0+m/\tau_1} \int \eu{-\tilde{S}(x_+,0,\tau)/\hbar} \, \delta(\sigma_\tau) \, \rmd x_+.
\end{align}
Using the properties of a delta function and performing the integral over $x_+$ gives
\begin{align}
	G_\tau(\gam{0},\gam{1})
	&= \sqrt\frac{2\pi\tau_0\tau_1}{\beta m} \frac{1}{\beta|\kappa_0-\kappa_1|} \,
\nonumber\\&\quad \times \eu{
			-\frac{\chi_{\tau}}{\hbar}
			\left[ \frac{\kappa_0^2\tau_0^2-\kappa_1^2\tau_1^2}{12m} - \mathbf{b_0} \cdot \mathbf{\gamma_0} + \mathbf{b_1} \cdot \mathbf{\gamma_1} \right]
			+ \frac{\kappa_0^2\tau_0^3 + \kappa_1^2\tau_1^3}{24m\hbar}
		},
\end{align}
where $\chi_{\tau}=\frac{\kappa_0 \tau_0 + \kappa_1\tau_1}{(\kappa_0-\kappa_1)}$. Using the fact that \cite{Feynman}
\begin{align}
	\lim_{N_n\rightarrow\infty}
	J_n \Lambda_n^{-N_n} \left|\frac{\mathbf{A_n}}{2\pi\hbar}\right|^{-\half} = \sqrt\frac{m}{2\pi\hbar\tau_n},
\end{align}
which we know by comparison with the exact path-integral expression for a linear potential, \cite{Brown1994linear}
performing the remaining Gaussian integrals gives
\begin{align}
	\label{ephix}
	\eu{-\phi(\tau)/\hbar}
	&= \sqrt\frac{m}{2\pi\beta\hbar^2} \frac{1}{\beta|\kappa_0-\kappa_1|} \,
		\eu{-\tilde{S}(\tau)/\hbar},
\end{align}
where
\begin{align}
	\tilde{S}(\tau) &=
		\chi_{\tau} \frac{\kappa_0^2\tau_0^2-\kappa_1^2\tau_1^2}{12m}
		- \frac{\kappa_0^2\tau_0^3 + \kappa_1^2\tau_1^3}{24m}
		\nonumber\\&\quad- \frac{\chi_{\tau}^2}{2}
		\left[ \mathbf{b_0} \cdot \mathbf{A_0}^{-1} \cdot \mathbf{b_0}
		+ \mathbf{b_1} \cdot \mathbf{A_1}^{-1} \cdot \mathbf{b_1} \right].
\end{align}

In the limit of an infinite number of modes, %
\begin{align}
\lim_{N_n\rightarrow\infty}
	\mathbf{b_n} \cdot \mathbf{A_n}^{-1} \cdot \mathbf{b_n}
	&= \sum_{k=1}^{\infty} \frac{2\kappa_n^2 (1-(-1)^k)^2 \tau_n}{m \pi^4 k^4}
	\\
	&= \frac{\kappa_n^2 \tau_n}{12 m},
\end{align}
and we find that actually $\tilde{S}(\tau)$ is independent of $\tau$ and is equal to the instanton action $\Sinst$, \eqn{Sinst}.
Therefore the GR-QTST rate will not depend on the choice of $\tau$,
and will be given by
\begin{align}
	k Z_0 = \sqrt\frac{2\pi m}{\beta\hbar^2} \frac{\Delta^2}{\hbar|\kappa_0-\kappa_1|} \, \eu{-\Sinst/\hbar},
\end{align}
which is also the instanton and exact Fermi's golden-rule result for a system of two linear potentials. \cite{GoldenGreens}
In the classical limit, the prefactor is the same, but the exponent disappears as $\Sinst\to0$.

Interestingly if we had ignored the fluctuation terms, GR-QTST would not have given the correct result.
This shows that it is not just the instanton, but also the fluctuations around all the constrained minima which contribute
to the path integral and to the success of the new method.

This proof is valid not just for systems with $\kappa_0\kappa_1<0$,
but also for those with $\kappa_0\kappa_1>0$,
which describe the Marcus inverted regime.
Note that the ring-polymer instanton method is not stable numerically in the inverted regime, whereas GR-QTST is.  

Although the reactant partition function, $Z_0$, is not well defined for this system,
this important model system describes a limiting case of many nonadiabatic reactions,
for which the reactant partition function can be defined.
Our findings here that GR-QTST gives excellent predictions for the rate constant are expected
to carry over at least approximately to more general systems.

\section{Analytic solution for an uncoupled linear and harmonic system}
\label{sec:linearharmonic}
In this section, we consider a multidimensional system formed of crossed linear potentials uncoupled to harmonic oscillators.
We choose it to be uncoupled such that the analytical result can be obtained.
The numerical result of a coupled problem is given in \secref{sysbath}.

For an uncoupled system, we have $V_n(x,y) = V_n^{(x)}(x) + V^{(y)}(y)$,
where $V_n^{(x)}(x)$ is linear as defined in  Appendix \ref{sec:linear} and $V^{(y)}(y) = \thalf m\omega^2y^2$.
For this system, the effective action is separable and can be written as a sum of the linear and harmonic parts,
$\phi(\tau) = \phi^{(x)}(\tau) + \phi^{(y)}(\tau)$,
where $\phi^{(x)}(\tau)$ is defined as in \eqn{ephix}. 

Here we just consider the effective action of the $y$-part, which is defined as follows: 
\begin{align}
	\eu{-\phi^{(y)}(\tau)/\hbar}
		&= J_0 \Lambda_0^{-N_0} J_1 \Lambda_1^{-N_1} \iint F_\tau(\mathbf{\lambda_0},\mathbf{\lambda_1}) 
		\nonumber\\&\quad
		\times G_\tau(\mathbf{\lambda_0},\mathbf{\lambda_1}) \, \rmd \mathbf{\lambda_0} \, \rmd \mathbf{\lambda_1},
\end{align}
where 
\begin{align}
	F_\tau(\mathbf{\lambda_0},\mathbf{\lambda_1}) = \exp\left[- \frac{1}{2\hbar}\mathbf{\lambda_0}\cdot\mathbf{B_0}\cdot\mathbf{\lambda_0} - \frac{1}{2\hbar}\mathbf{\lambda_1}\cdot\mathbf{B_1}\cdot\mathbf{\lambda_1} \right]
\end{align}
and
\begin{align}
G_\tau(\mathbf{\lambda_0},\mathbf{\lambda_1})&=\iint \eu{-\tilde{S}^{(y)}(\tau)/\hbar}
\nonumber\\&\quad
\times \eu{\chi_{\tau} [\sigma^{(y)}(\tau)-(\frac{1}{2}c\mathbf{\lambda_1}^2-\frac{1}{2}c\mathbf{\lambda_0}^2)]/\hbar} \, \rmd y' \, \rmd y'',
\end{align}
The action in the $y$ degree of freedom is $\tilde{S}^{(y)}(\tau)=\tilde{S}^{(y)}_{0}(\tau_0)+\tilde{S}^{(y)}_{1}(\tau_1)$, where
\begin{align}
\tilde{S}^{(y)}_n(\tau_n)=\frac{m \omega}{2 \sinh(\omega \tau_n)}\left((y'^{2}+y''^{2}) \cosh(\omega \tau_n)-2 y' y''\right).
\end{align}

For a harmonic oscillator, the fluctuations in the classical action are given in terms of the Fourier coefficients
$\mathbf{\lambda_n}=(\lambda_1^{(n)},\dots,\lambda_{N_n-1}^{(n)})$ as
\begin{align}
	\thalf \mathbf{\lambda_n}\cdot\mathbf{B_n}\cdot\mathbf{\lambda_n} &= \sum_{k=1}^{N_n-1} \frac{m (\pi k \lambda_k^{(n)})^2}{4\tau_n}
	+ \tfrac{1}{4}m \omega^2 \tau_n (\lambda_k^{(n)})^2.
\end{align} 
Likewise the fluctuations in energy are given as
\begin{align}
	\tfrac{1}{2}c\mathbf{\lambda_n}^2 = \sum_{k=1}^{N_n-1} \tfrac{1}{2}m \omega^2 (\lambda_k^{(n)})^2,
\end{align}
where $c=m \omega^2$, which is just a constant independent of index $k$.

The constraint functional for the uncoupled linear and harmonic potential-energy surface is the sum of the energy-matching functionals in $x$ and $y$ directions,
\begin{align}
	\sigma^{(x,y)}(\tau)=\sigma^{(x)}(\tau)+\sigma^{(y)}(\tau). 
\end{align}
where $\sigma^{(x)}(\tau)$ is the constraint functional for the one-dimensional problem, \eqn{constraint}.
For the end points of the classical path, the energy matching function for the harmonic oscillator simply is
\begin{align}
	\sigma^{(y)}(\tau) &= \tfrac{2}{3}\beta \big(\eta^{(y)}_0(\tau_0)-\eta^{(y)}_1(\tau_1) \big),
\end{align}
where 
\begin{align}
\eta^{(y)}_n(\tau_n)=\frac{1}{\tau_n} \int_0^{\tau_n} \tfrac{3}{2} m \omega^2 y^2 \, \rmd t,
\end{align}
which is the classical energy corresponding to one of the trajectories projected onto the $y$-direction.

First we perform Gaussian integrals over fluctuation variables, $\mathbf{\lambda_n}$, yielding
\begin{align}
\label{ephiy}
\eu{-\phi^{(y)}(\tau)/\hbar}=&\frac{m}{2 \pi \hbar \tau_0 \tau_1} \sqrt{\psi_0\psi_1} \iint \eu{-\tilde{S}^{(y)}(\tau)/\hbar}
\nonumber\\&\quad\times \eu{\chi_{\tau} \sigma^{(y)}(\tau)/\hbar} \, \rmd y' \, \rmd y'',
\end{align}
where
\begin{align}
\psi_n = \frac{\omega \sqrt{\tau_n(\tau_n-2 \chi_{\tau})}}{\sinh(\omega \sqrt{\tau_n(\tau_n-2 \chi_{\tau})}}
.
\end{align}
Here we have used the following identities given by Feynman \cite{Feynman}
\begin{align}
\label{yprefac}
\begin{split}
	\lim_{N_0\rightarrow\infty}
	J_0 \Lambda_0^{-N_0} |\mathbf{B_0}/(2\pi\hbar)|^{-\half} = \sqrt\frac{m \omega}{2\pi\hbar \sinh\omega \tau_0}
\\
	\lim_{N_1\rightarrow\infty}
	J_1 \Lambda_1^{-N_1} |\mathbf{B_1}/(2\pi\hbar)|^{-\half} = \sqrt\frac{m \omega}{2\pi\hbar \sinh\omega \tau_1}.
\end{split}
\end{align}

The integral over end points, $y'$ and $y''$, are also Gaussian, and yields the following result:
\begin{equation}
\iint \eu{-\tilde{S}^{(y)}(\tau)/\hbar+\chi_{\tau} \sigma^{(y)}(\tau)/\hbar} \, \rmd y' \, \rmd y''= \frac{2 \pi}{\sqrt{\det\mathbf{C}}},
\end{equation}
where the diagonal entries of the 2 $\times$ 2 matrix $\mathbf{C}$ are
\begin{equation}
	c_{11}=c_{22}=m \omega\left[\coth (\omega \tau_0)+\coth (\omega \tau_1)-\chi_{\tau}(\alpha_0-\alpha_1)\right],
\end{equation}
where $\alpha_n=\coth (\omega \tau_n)/\tau_n-\omega \csch^2(\omega \tau_n)$.
The off-diagonal entries are
\begin{equation}
	c_{12}=c_{21}=-m \omega \left[\csch (\omega \tau_0)+\csch (\omega \tau_1)+ \chi_{\tau}(\theta_0-\theta_1)\right], 
\end{equation}
where $\theta_n = \csch (\omega \tau_n) \left(\omega \coth (\omega \tau_n)-1/\tau_n\right)$.
Therefore the effective action is defined by
\begin{align}
\eu{-\phi^{(y)}(\tau)/\hbar}=  \frac{m }{\hbar \tau_0 \tau_1} \sqrt\frac{\psi_0\psi_1}{\det\mathbf{C}}.
\end{align}

Under the GR-QTST formulation, the effective action is evaluated at $\tau=\tau^{*}$, which corresponds to the point where $\phi^{(y)}(\tau)$ demonstrates the maximum.
For this system, this maximum occurs at $\tau^*=\tauinst$ for which
$\chi_{\tauinst} = 0$.
Therefore the effective action simplifies to
\begin{align}
\eu{-\phi_y(\tau^*)/\hbar}=\frac{1}{2 \sinh(\beta \hbar \omega/2)}.
\end{align}

This quantity is recognized as the partition function in the $y$-direction, which cancels with an equivalent term in the reactant partition function.
Therefore the GR-QTST rate obtained for an uncoupled linear and harmonic system is exactly the same as the rate obtained for a one-dimensional linear system, which is of course correct.
This derivation also holds for many uncoupled oscillators
and remains at least approximately true for a coupled bath (see \fig{phi}).

\clubpenalty10000 %

%
%
%

%
%
%
%

\end{document}